\documentclass[showpacs,aps,prb,twocolumn,superscriptaddress,preprintnumbers,amsmath,amssymb,floatfix,10pt]{revtex4}
\usepackage{graphicx}% Include figure files
\usepackage[tight]{subfigure}
\usepackage{dcolumn}% Align table columns on decimal point
\usepackage{float}
\usepackage{bm}% bold math
\usepackage{enumerate}
\usepackage[english]{babel}
\usepackage{amsfonts}
\usepackage{amsthm}
\usepackage{amssymb}
\usepackage{calc}
\usepackage{color}
\usepackage{ifthen}
\usepackage{hyperref}
\newcommand{\bfw}{\textbf{w}}
\newcommand{\bfz}{\textbf{z}}

\newcommand{\bfr}{\textbf{r}}

\newcommand{\bfq}{\textbf{q}}
\newcommand{\bfM}{\textbf{M}}

\newcommand{\bsT}{\boldsymbol{\Theta}}
\newcommand{\bsO}{\boldsymbol{\Omega}}
\newcommand{\e}{\mathrm{e}}
\usepackage{inputenc}
\inputencoding{utf8}

\begin{document}
\title{Screening properties and phase transitions in unconventional plasmas for Ising-type quantum Hall states}

\author{Egil V. Herland}
\affiliation{Department of Physics, Norwegian University of Science and Technology, N-7491 Trondheim, Norway}

\author{Egor Babaev}
\affiliation{Physics Department, University of Massachusetts, Amherst, Massachusetts 01003, USA}
\affiliation{Department of Theoretical Physics, The Royal Institute of Technology, 10691 Stockholm, Sweden}

\author{Parsa Bonderson}
\affiliation{Station Q, Microsoft Research, Santa Barbara, California 93106-6105, USA}

\author{Victor Gurarie}
\affiliation{Department of Physics, CB 390, University of Colorado, Boulder, Colorado 80309, USA}

\author{Chetan Nayak}
\affiliation{Station Q, Microsoft Research, Santa Barbara, California 93106-6105, USA}
\affiliation{Department of Physics, University of California, Santa Barbara, California 93106, USA}

\author{Asle Sudb\o}
\affiliation{Department of Physics, Norwegian University of Science and Technology, N-7491 Trondheim, Norway}

\begin{abstract}
Utilizing large-scale Monte-Carlo simulations, we investigate an
unconventional two-component classical plasma in two dimensions
which controls the behavior of the norms and overlaps of the quantum-mechanical
wavefunctions of Ising-type quantum Hall states.
The plasma differs fundamentally from that which is associated with
the two-dimensional XY model and Abelian fractional quantum Hall states.
We find that this unconventional plasma undergoes a Berezinskii-Kosterlitz-Thouless
phase transition from an insulator to a metal. The parameter values
corresponding to Ising-type quantum Hall states
lie on the metallic side of this transition. This result verifies the required properties
of the unconventional plasma used to demonstrate that Ising-type 
quantum Hall states possess quasiparticles with non-Abelian braiding statistics.
\end{abstract}

\pacs{73.43.Cd, 74.20.-z}

\maketitle

\section{Introduction}
Key properties of physical systems can sometimes be understood by mapping them to seemingly unrelated ones.
A powerful example of this was provided by Laughlin, who observed that the squared norm of his $\nu=1/M$ fractional quantum Hall trial wavefunction
\begin{equation}
{\Psi}(z_i) =  \prod_{i<j}^{N} (z_i-z_j)^M \, \mathrm{e}^{- \frac 1 4 \sum\limits_{i=1}^N \left| z_i \right|^2 }
\end{equation}
(where $z_i = x_i + i y_i$ is a complex coordinate in the two-dimensional plane)
could be expressed as the Boltzmann weight of a two-dimensional
one-component plasma~\cite{Laughlin1983}:
\begin{equation}
\left\| \Psi (z_i) \right\|^2 =
\int \prod_{i=1}^{N} {\textrm{d}^2}z_i \, |{\Psi}(z_i)|^2 = \int \prod_{i=1}^{N} {\textrm{d}^2}z_i \,\mathrm{e}^{-\beta {V_1}(z_i)}
\end{equation}
where
\begin{equation}
\label{eqn:Laughlin-plasma}
{V_1}(z_i) = - {Q_1^2} \sum_{i<j}^N \ln \left| z_i - z_j \right|
 +  \frac{Q_1^2}{4M} \sum_{i=1}^N \left| z_i \right|^2
\end{equation}
and ${Q_1^2}/T = 2M$.
This mapping allows properties such as quasiparticle charge and braiding statistics to be determined by appealing to the known properties of a one-component plasma.

Recently, a similar plasma mapping was
established~\cite{Bonderson_PRB_2011} for Ising-type quantum Hall
states, such as the Moore-Read (MR)~\cite{Read1991},
anti-Pfaffian~\cite{Lee07,Levin07}, and Bonderson-Slingerland (BS)
hierarchy~\cite{Bonderson08} states, which are likely candidates to
describe Hall plateaus in the second Landau level, in particular at
filling fraction
$\nu=5/2$~\cite{Willett87,Pan99b,Eisenstein02,Xia04}. In this case,
the mapping is to a two-dimensional (2D) two-component plasma, where
the two species of particles, $w$ and $z$, carry not only different
values of charge, but also interact through two different interactions,
both of the Coulomb form, so the potential energy is:
\begin{equation}
\label{eqn:two-Coulomb-ints}
V(z_i; w_a) = V_1(z_i) + V_2(z_i; w_a),
\end{equation}
\begin{align}
\label{eqn:two-comp-plasma}
V_2(z_i; w_a) = &- Q_2^2 \sum_{i<j}^N \ln \left| z_i - z_j \right| -
Q_2^2 \sum_{a<b}^N \ln \left| w_a -w_b \right|\nonumber\\ &+ Q_2^2 \sum_{a,i}^N \ln \left| z_i - w_a \right|,
\end{align}
where ${Q_2^2}/T=3$. The $z$-particles interact with each other
through the first Coulomb-like interaction, $V_1(z_i)$, given in Eq.~\eqref{eqn:Laughlin-plasma} (and so does not depend on the $w_a$
coordinates). Moreover, the $z$-particles interact with each other and with the $w$-particles
through the second Coulomb-like interaction, through which the
$w$-particles also interact with each other, according to
$V_2(z_i;w_a)$, given in Eq.~\eqref{eqn:two-comp-plasma}. Note that $V_2(z_i;w_a)$ is the 2D Coulomb potential of the
usual two-component plasma (where the two species carry charge $Q_2$ and
$-Q_2$, respectively).

\begin{figure}[tbp]%%[htbp]
\includegraphics[width=\columnwidth]{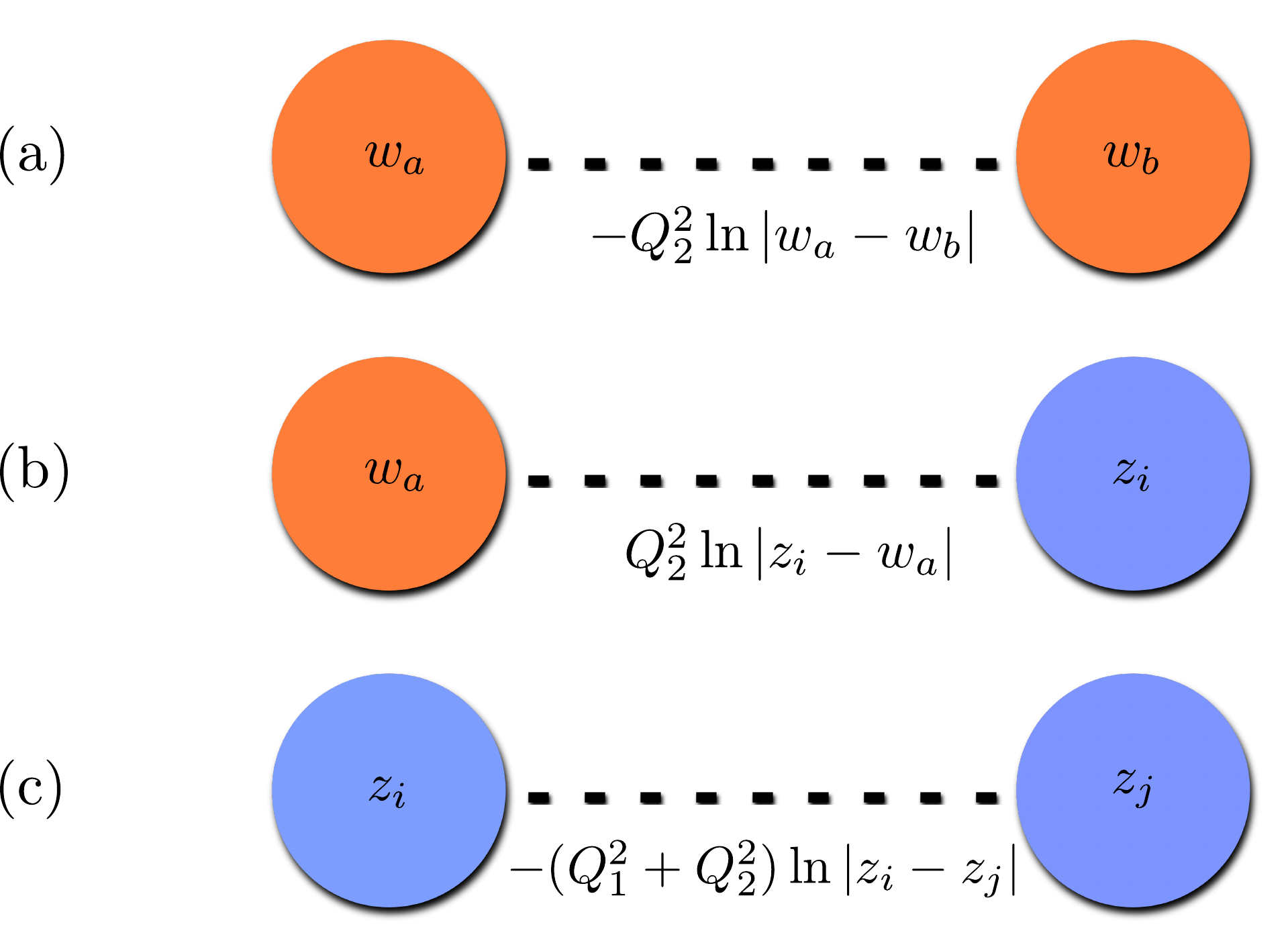}
\caption{(Color online) Illustration of interactions between the
  particles in the 2D system. The $w$-particles only interact by
  the second Coulomb-like interaction with charge $-Q_2$, whereas the
  $z$-particles carry charge $Q_1$ for the first Coulomb-like
  interaction and $Q_2$ for the second Coulomb-like interaction. Thus,
  the intraspecies interaction among the
  $w$-particles, shown in (a), and the interspecies interaction between $w$ and
  $z$-particles, shown in (b), are given by $Q_2$ only, whereas the intraspecies interaction among the
  $z$-particles, shown in (c), are determined by $Q_1$ in addition to $Q_2$. Interactions between the $z$-particles and the neutralizing background are omitted
  from the figure.}
\label{Fig:Interactions}
\end{figure}

The $z$-particles carry charge $Q_1$ for the first interaction
and charge $Q_2$ for the second interaction.
The $w$-particles carry charge $0$ for the first interaction
and charge $-Q_2$ for the second interaction.
For a plasma with $N$ particles of each species, neutrality is
satisfied using a uniform background density of type 1 charge,
as in the second term in Eq.~\eqref{eqn:Laughlin-plasma}.
This unconventional plasma may be considered as an ordinary neutral
two-component gas with positive and negative charges of
magnitude $Q_2$, where the positive charges are given an additional
charge of $Q_1$ that is only felt by the other positive charges and
not the negative charges. An illustration of the interactions between
the two species in the system is shown in Fig.~\ref{Fig:Interactions}.

We are thus led to consider a class of unconventional plasmas parametrized
by $Q_1^2 / T$ and $Q_2^2 / T$. As mentioned above,
for MR Ising-type states with filling $\nu=1/M$, the relevant values
are $Q_1^2 / T = 2M$ and $Q_2^2 / T = 3$. In this plasma mapping, the $z_i$ particles in the plasma correspond to the electrons in the MR wavefunctions and the $w_a$ particles correspond to screening operators (fictitious particles).
The case $Q_1 =0$, $Q_2^2 / T = 3$ is relevant for the plasma mapping~\cite{Bonderson_PRB_2011} of 2D chiral $p$-wave superconductors~\cite{Read2000}.
 We note that whenever $Q_1 =0$, our model is a special case of
 the well-known 2D two-component plasma of equal and opposite charges~\cite{Kosterlitz_JPhysC_1973, Kosterlitz_JPhysC_1974,Caillol_PRB_1986, Orkoulas_JChemPhys_1996}.
 The screening properties of multi-component 2D plasmas with multiple Coulomb interactions of this kind are also important for other physical systems, such as rotating multi-component Bose-Einstein condensates with interspecies
current-current (Andreev-Bashkin) interaction~\cite{Dahl_PRB_2008,Dahl_PRL_2008} and
some multi-component superconducting systems~\cite{Smiseth_PRB_2005, Herland_PRB_2010, Babaev_PRB_2008}. In these systems the screening properties and phase transitions determine superfluid and rotational responses.

In this paper, we fix temperature to $T=1$ and consider the two most
significant values of $Q_1$, namely $Q_1 = 0, 2$. We investigate the screening and phase transition properties of these plasmas as a function of varying $Q_2$ by performing a large-scale Monte Carlo
simulation. Here, a "screening phase" means that the system has a screening length which is finite,
and exponentially decaying effective interactions. A system with
logarithmic effective interactions is one where screening is defined to be absent.
As a first check, we reproduce the well-known result that, for $Q_1 = 0$, there
is a Berezinskii-Kosterlitz-Thouless (BKT) phase transition at $Q_2^2 = Q_{2,\text{c}}^2 \approx 4$,
as expected for a 2D two-component plasma of equal and opposite charges.
For $Q_2^2 < Q_{2,\text{c}}^2$, the charges are unbound and the
plasma screens, but for $Q_2^2 > Q_{2,\text{c}}^2$, the charges are bound into
dipoles and the interaction is not screened. Thus, for $Q_2^2 = 3$, the
value relevant to 2D chiral $p$-wave superconductors, the plasma screens.
For $Q_1 = 2$, we again find a BKT phase transition at $Q_2^2 = Q_{2,\text{c}}^2 \approx 4$,
with a plasma screening phase for $Q_2^2 < Q_{2,\text{c}}^2$. The first Coulomb-like interaction
is deep within its screening phase and appears to have a negligibly small effect on the
screening of the second interaction. In both cases, the critical values $Q_{2,\text{c}}^2$
are obtained by a finite-size scaling fit of the Monte Carlo data to
the BKT form. Our findings demonstrate that the unconventional plasma
which occurs in the mapping for both a chiral $p$-wave superconductor and
the Ising-type quantum Hall states is clearly in the screening phase (for
both types of Coulomb interaction) and hence allows one to discern the
non-Abelian braiding properties of these states, as explained in
Ref.~\onlinecite{Bonderson_PRB_2011}.

The outline of this paper is as follows. In the introductory part of Section~\ref{sec:Model}, we present the model for
the unconventional plasma we will be studying in this paper. In Section~\ref{sec:Ising_type_QHS}, we connect this to the
Ising-type of quantum Hall states. In Section~\ref{sec:rotating_BECs}, we explain its connection to two-component, two
dimensional, Bose-Einstein condensates. In Section~\ref{sec:spherical}, we present a formulation of the model on a
sphere. In Section~\ref{sec:MC_details}, we give details of the Monte-Carlo simulations, and in Section~\ref{sec:results}, we present
our results for the screening properties, as well as our findings for the character of phase transition
between the dielectric non-screening phase and the metallic screening phase. In Section~\ref{sec:conclusions}, we present
our conclusions. Technical details on the derivation of a generalized dielectric constant is given in
Appendix~\ref{App:General_dielectric_constant}. In Appendix~\ref{App:Fourth_order_derivative}, we give a derivation of a relevant higher order response function that we use to characterize the metal-insulator transition. In Appendix~\ref{App:Weber_Minnhagen}, we present technical details on the finite-size scaling we have used.

\section{Model}
\label{sec:Model}

The canonical partition function of the unconventional plasma is written
\begin{equation}
Z = \int \left(\prod_{i=1}^N \textrm{d}^2z_i \right) \left(\prod_{a=1}^N \textrm{d}^2w_a \right)\mathrm{e}^{-V},
\label{Eq:Model_Z}
\end{equation}
where the potential energy $V$ is given by the 2D Coulombic interactions
\begin{align}
V = & \quad Q_2^2 \sum_{a<b=1}^N v_{ww}(|\bfw_a-\bfw_b|)\nonumber\\
&+ (Q_1^2 + Q_2^2)\sum_{i<j=1}^N v_{zz}(|\bfz_i-\bfz_j|)\nonumber\\
&+ Q_2^2 \sum_{a,i=1}^N v_{zw}(|\bfz_i-\bfw_a|) + V_{z, \textnormal{BG}} .
\label{Eq:Model_S}
\end{align}
Similar to the study of the 2D two-component neutral Coulomb gas~\cite{Kosterlitz_JPhysC_1973, Kosterlitz_JPhysC_1974,Caillol_PRB_1986, Orkoulas_JChemPhys_1996, Note_1}, we introduce a short-range hard-core repulsion between all charges in the system. Treating all charges as hard disks with the same diameter $d$ that limits the range of the hard-core repulsion, the interaction between charges of the same species is
\begin{equation}
v_{zz}(|\bfr|) = v_{ww}(|\bfr|) = \left\{ \begin{array}{lr}
\infty, & |\bfr| \leq d,\\
-\ln|\bfr|, & |\bfr| > d,\\
\end{array} \right.
\label{Eq:Same_component_interaction}
\end{equation}
and the interaction between charges of different species is
\begin{equation}
v_{zw}(|\bfr|) = \left\{ \begin{array}{lr}
\infty, & |\bfr| \leq d,\\
\ln|\bfr|, & |\bfr| > d.\\
\end{array} \right.
\label{Eq:Different_component_interaction}
\end{equation}

In Eq.~\eqref{Eq:Model_S}, $\bfw_a$ are position vectors for the particles of component $w$, and $\bfz_i$
are position vectors for the particles of component $z$. To ensure
neutrality, the term $V_{z, \textnormal{BG}}$ includes the interaction of the $Q_1$
charges of type 1 for the $z$-particles with a neutralizing background charge density. In
Ref.~\onlinecite{Bonderson_PRB_2011}, this background is a uniform
negatively charged 2D disk with charge density $q^{\textnormal{BG}}_{1} = - N Q_1 / A$,
where $N/A = 1/2\pi M$, that yields
\begin{equation}
V_{z, \textnormal{BG}} = \frac{1}{2}\sum_{i=1}^N \left| z_i \right|^2.
\label{Eq:BZ_term}
\end{equation}
The particle-background and the background-background interaction also
yields uninteresting constant terms, that
are disregarded in Eq.~\eqref{Eq:Model_S}.

We note that when $Q_1 = 0$ we have the 2D
two-component neutral Coulomb plasma, which is well-studied both
analytically and numerically~\cite{Kosterlitz_JPhysC_1973,
  Kosterlitz_JPhysC_1974, Caillol_PRB_1986, Berezinskii_JETP_1971,
  Minnhagen_RevModPhys_1987, Saito_PRB_1981,
  Lee_PRB_1992, Lidmar_PRB_1997, Gupta_PRB_1997,
  Orkoulas_JChemPhys_1996}. At low dipole density, this system will
undergo a BKT transition, which is a
charge-unbinding transition from a
low-temperature state where charges of opposite signs form tightly bound dipoles to
a high-temperature state in which a finite fraction of charges are
not bound in dipoles, but rather form a metallic state. In the low-temperature phase,
this Coulomb gas is an insulator and the dielectric constant
$\epsilon$ (see for instance
Refs.~\onlinecite{Minnhagen_RevModPhys_1987, Olsson_PRB_1992,
  Olsson_PRB_1995} and Appendix~\ref{App:General_dielectric_constant}
for a formal definition of $\epsilon$), is finite. In the
high-temperature phase, the existence of free charges, yields a
conductive gas with an infinite value of $\epsilon$. At the
critical temperature $T_{\text{c}}$, when tightly bound dipoles starts to unbind,
there is a universal jump in the inverse dielectric constant from a
non-zero value in the insulating phase to zero in the metallic phase,
\begin{equation}
\epsilon^{-1} = \left\{ \begin{array}{lr}
4 T_{\text{c}}, & T \rightarrow T_{\text{c}}^-,\\
0, & T \rightarrow T_{\text{c}}^+.\\
\end{array} \right.
\label{Eq:Universal_jump}
\end{equation}
The screening properties that follow are that the
Coulomb gas is able to perfectly screen test charges in the metallic
phase when there are free charges in the system, whereas there is no
screening in the insulating dielectric phase. In this work, we will
focus our attention on the low dipole density regime, so we will not go
into details of the physics in the 2D two-component neutral
Coulomb gas at higher densities. However, we note that when
density is increased, the critical point of the BKT transition is shifted
towards lower temperatures~\cite{Caillol_PRB_1986, Lee_PRB_1992,
  Lidmar_PRB_1997, Orkoulas_JChemPhys_1996}.

Another well-studied case is when $Q_2 = 0$, for which the model
reduces to the 2D one-component plasma (for the $z$-particles only). Early numerical studies of this system found a weak
first-order melting transition at $Q_1^2 / T \approx 140$ from a state where
the charges form a triangular lattice with quasi-long-range translational and
long-range orientational order to a fluid plasma state~\cite{de_Leeuw_PhysicaA_1982,
  Caillol_JStatPhys_1982, Choquard_PRL_1983, Franz_PRL_1994}. These
results were, in a sense, contrasting with the defect-mediated melting
theory of Kosterlitz-Thouless-Halperin-Nelson-Young (KTHNY) that predicts melting from a solid to a liquid via two
BKT-transitions and an intermediate hexatic phase with no
translational order and quasi-long-range orientational
order~\cite{Kosterlitz_JPhysC_1973, Halperin_PRL_1978,
  Nelson_PRB_1979, Young_PRB_1979}. Other studies of 2D melting point
in favor of the KTHNY theory~\cite{Chen_PRL_1995,
  Perez-Garrido_PRB_1998, Dietel_PRB_2006, Lee_PRE_2008}, suggesting that the nature of melting
transition may depend on details in the interatomic potential, or that
finite-size effects and lack of equilibration
might lead to erroneous conclusions in earlier works. There are also studies
that argue for the absence of a phase transition to a low-temperature
solid phase in the 2D one-component plasma with repulsive logarithmic
interactions because the crystalline state would be unstable to
proliferation of screened disclinations for any $T > 0$~\cite{O'Neill_PRB_1993, Dodgson_PRB_1997, Moore_PRL_1999, McClarty_PRB_2007}.

\subsection{Ising-Type quantum Hall states}
\label{sec:Ising_type_QHS}

The unconventional 2D two-component plasma studied here is mapped to inner products of trial wavefunctions for the MR quantum Hall states using conformal field theory (CFT) methods, as explained in Ref.~\onlinecite{Bonderson_PRB_2011}. In particular, this mapping utilizes the Coulomb gas description of CFTs~\cite{Dotsenko1984,Felder1989} together with a procedure for replacing holomorphic-antiholomorphic pairs of contour integrals in screening charge operators for 2D integrals~\cite{Mathur1992,Bonderson_PRB_2011}.

The MR states' wavefunctions can be written as a product of correlation functions of fields from the Ising and U(1) CFTs.
In particular, the MR ground-state wavefunction for $N$ electrons is
\begin{equation}
\label{eq:pf}
\Psi \left(z_1,\dots, z_{N} \right)  =
{\rm Pf}\left(\frac{1}{z_i-z_j}\right)\,\prod_{i<j}^{N}
\left( z_i-z_j \right)^M \mathrm{e}^{  - \frac{1}{4} \sum\limits_{i=1}^{N}  \left| z_i \right|^2 }
\end{equation}
where the Pfaffian of an antisymmetric matrix $A$ is given by
\begin{equation}
\label{eq:pfaffian}
{\rm Pf}\left(A_{i,j}\right) \equiv \frac{1}{N!!} \sum_{\sigma \in S_{N}} \text{sgn} (\sigma) \prod_{k=1}^{N/2} A_{ \sigma (2k-1), \sigma (2k)}
.
\end{equation}
Here, $S_N$ is the symmetric group, $\sigma$ is one of the permutation elements in $S_N$, and $\rm{sgn}(\sigma)$
is the signature of $\sigma$. The ${\rm Pf}\left(\frac{1}{z_i-z_j}\right)$ portion of this wavefunction is produced 
from the correlation function of $\psi$ fields in the Ising CFT, while the Laughlin-type portion
\begin{equation}
\prod_{i<j}^{N} \left( z_i-z_j \right)^M \mathrm{e}^{  - \frac{1}{4} \sum\limits_{i=1}^{N}  \left| z_i \right|^2 }
\end{equation}
is produced from the correlation function of vertex operators in the U(1) CFT.

The Laughlin-type portion of the MR wavefunctions can be mapped to charges of type 1, similar to Laughlin's plasma mapping.
The mentioned CFT methods provide identities such as
\begin{align}
\label{eq:pf_map}
\left| {\rm Pf}\left(\frac{1}{z_i-z_j}\right) \right|^2 =& \int \prod_{a=1}^{N} \textrm{d}^{2} w_{a} \prod_{a<b}^{N} \left| w_a - w_b \right|^{3} \notag \\
&\times \prod_{i<j}^{N} \left| z_i - z_j \right|^{3} \prod_{a,i}^{N} \left| w_a - z_i \right|^{-3}
,
\end{align}
which allow the Pfaffian portion of the MR wavefunctions to be mapped to charges of type 2. This allows one to write the
norm of the MR ground-state wavefunction as the partition function of the unconventional 2D two-component plasma of
Eq.~(\ref{eqn:two-Coulomb-ints})
\begin{eqnarray}
\label{eq:norm}
\left\| \Psi \left(z_1,\dots, z_{N} \right) \right\|^2 &=& \int \prod_{i=1}^{N} \textrm{d}^{2} z_{i} \left| \Psi \left(z_1,\dots, z_{N} \right) \right|^2 \notag \\
&=& \int \prod_{a=1}^{N} \textrm{d}^{2} w_{a} \prod_{i=1}^{N} \textrm{d}^{2} z_{i} \,\, \mathrm{e}^{-V}
,
\end{eqnarray}
with $Q_1^2 = 2M$ and $Q_2^2 = 3$. More generally, one can also construct a similar, but more complicated mapping between inner
products of wavefunctions of the MR states with quasiparticles, as explained in Ref.~\onlinecite{Bonderson_PRB_2011}. In this
case, the quasiparticles map to fixed ``test'' objects in the plasma that carry electric charge of type 1 and can carry both
electric and magnetic charges of type 2 (and also changes the number of screening operators, i.e. $w$-particles in the plasma,
to maintain neutrality). (The charges of type 1 and 2 carried by the quasiparticles are typically some fractions of the
charges $Q_1$ and $Q_2$ carried by the $z$-particles.)

Strictly speaking, the right-hand-side of Eq.~(\ref{eq:pf_map}) is divergent for $Q_2^2 =3$ (since the integrand diverges as
$\left| w_a - z_i \right|^{-3}$ as a $w$-particle approaches a $z$-particle). It can be made well-defined
(and equal to the left-hand-side) by replacing $\left| w_a - z_i \right|^{-3}$ with
$\left| w_a - z_i \right|^{-\alpha}$, evaluating the integrals for $\alpha<2$ and then
analytically-continuing $\alpha$ to $3$. On the other hand, we regularize the divergences of Eq.~(\ref{eq:norm}) in this paper by using
a hard-core repulsion that forbids the particles from approaching each
other closer than a distance $d$, i.e. replacing $V$ in this expression with that of Eq.~(\ref{Eq:Model_S}).
It should not matter how we regularize the divergence in Eq.~(\ref{eq:norm}) as long as the
probability for $z$-particles and $w$-particles to sit right on top of each other has measure zero.
As we will see in this paper, this is true for $Q_2^2 < Q_{2,\text{c}}^2 \approx 4$,
in which case the configurational entropy to be gained by having $z$-particles and $w$-particles separate
overcomes the energy gained by having them on top of each other. We refer to this as an ``entropic
barrier'' for putting $z$-particles and $w$-particles on top of each other.
In contrast, in Eq.~(\ref{eq:pf_map}), where only the
$w_i$s are integrated over and the $z_i$ coordinates are fixed,
regularization by a simple hard-core repulsion does not appear to be a suitable alternative to analytic continuation.
In this case, since the $z_i$ coordinates are fixed, the entropic barrier is lower. Equivalently,
there are fewer integrals to compensate the inverse powers. Thus, in Eq.~(\ref{eq:pf_map}),
a simple hard-core cutoff will not reproduce the left-hand-side, and one must use the analytic continuation
procedure described above.

\subsection{Two component rotating Bose-Einstein condensate in two dimensions}
\label{sec:rotating_BECs}

In a rotating frame, a Bose-Einstein condensate in the London limit is
described by the uniformly frustrated XY model,
\begin{equation}
H = \frac{\rho}{2}\int \mathrm{d}^2r\left[\nabla \theta (\bfr) -
\frac{m}{\hbar}\bsT (\bfr)\right]^2,
\label{Eq:Frustrated_3Dxy}
\end{equation}
where $\rho = \hbar^2 n/m$ for a condensate with mass $m$, phase
$\theta$ and density $n$, and $\bsT(\bfr) = \bsO \times \bfr$ where
$\bsO = \Omega \hat{z}$ is the angular velocity of the rotation. In 3D, this model is
frequently used to describe the melting of vortex-line lattices in
extreme type-II superconductors and superfluids~\cite{Hetzel_PRL_1992,
Ryu_PRB_1998, Olsson_PRB_2003, Kragset_PRL_2006}. By a
duality transformation, the model in Eq.~\eqref{Eq:Frustrated_3Dxy}
can be rewritten in terms of vortex fields $l$ to
yield~\cite{Fradkin_PRB_1978, Chen_PRB_1997}
\begin{align}
H = \frac{1}{2}\int \mathrm{d}^2q&\left[l(\bfq)-(2\pi)^2
  f\delta(\bfq)\right]\frac{\rho}{q^2}\nonumber\\
&\times\left[l(-\bfq)-(2\pi)^2f\delta(-\bfq)\right],
\label{Eq:Frustrated_3Dxy_dual}
\end{align}
where $f = 2\Omega/\phi_o$ is the vortex number density and $\phi_0 =
2\pi\hbar/m$ is the fundamental quantum unit of vorticity. This is
a one-component 2D classical Coulomb plasma where charges correspond to nonzero values in the vortex field $l(\bfr)$ and the quantity
$f$ now plays the role as the neutralizing background number density.

Extending to two components, a model for a rotating two-component
Bose-Einstein condensate with a generic Andreev-Bashkin drag
interaction~\cite{Andreev_JETP_1975,kuklov_PRL_2003,kuklov_PRL_2004} reads
\begin{align}
H =& \frac{1}{2} \int \mathrm{d}^2r \Bigg\{ \sum_{i = 1,2}m_i
 n_i\left(\frac{\hbar\nabla \theta_i}{m_i} - \bsT\right)^2\nonumber\\
&-\sqrt{m_1m_2}n_d\left ( \frac{\hbar\nabla \theta_1}{m_1}-\frac{\hbar\nabla \theta_2}{m_2} \right )^2\Bigg\},
\label{Eq:Rotating_2comp_BEC}
\end{align}
where now $m$, $n$ and $\theta$ is given an index that denotes the
component and $n_d$ is the drag density. This model has recently been
studied in three dimensions~\cite{Dahl_PRB_2008, Dahl_PRL_2008}.
By a duality transformation, we arrive at the following 2D Coulomb plasma
\begin{align}
H =& \frac{1}{2}\int \mathrm{d}^2q\left[l_i(\bfq)-(2\pi)^2f_i
  \delta(\bfq)\right]\frac{R_{ij}}{q^2}\nonumber\\
&\times\left[l_j(-\bfq)-(2\pi)^2f_j \delta(-\bfq)\right],
\label{Eq:Rotating_2comp_BEC_dual}
\end{align}
where $f_i = 2\Omega/\phi_{0,i}$, $\phi_{0,i} = 2\pi\hbar/m_i$, $l_i$
is the vortex field of component $i$,
\begin{equation}
R = \hbar^2 \begin{pmatrix} \dfrac{1}{m_1}\left(n_1-\sqrt{\dfrac{m_2}{m_1}}n_d\right) & \dfrac{1}{\sqrt{m_1m_2}}n_d \\
\dfrac{1}{\sqrt{m_1m_2}}n_d & \dfrac{1}{m_2}\left(n_2-\sqrt{\dfrac{m_1}{m_2}}n_d\right)
\end{pmatrix},
\label{Eq:R_matrix}
\end{equation}
and an implicit sum over repeated component indices $i$, $j$ is
assumed. By setting $\hbar = m_i = 1$ such that $f_1 = f_2 = f$, and
absorbing a factor $2\pi\beta$ in the
density coefficients, we see that the two-component Bose-Einstein
condensate in Eq.~\eqref{Eq:Rotating_2comp_BEC} with $n_1 = 0$, $n_2 =
Q_1^2$ and $n_d = -Q_2^2$ corresponds to the
unconventional two-component Coulomb plasma in
Eq.~\eqref{Eq:Model_S}. Thus, the unconventional Coulomb plasma has a
counterpart in a two-component Bose-Einstein condensate with a
negative non-dissipative drag interaction. However, note that in order to
preserve a fixed number of charges when going from the plasma description in
Eq.~\eqref{Eq:Model_S} to the phase description in Eq.~\eqref{Eq:Rotating_2comp_BEC}, we have to fix
the number of vortices to only include rotationally induced vortices.
In principle, in the BEC problem, the system
can thermally excite vortex-antivortex pairs, but that process can be substantially suppressed
by going beyond the phase only model in Eq.~\eqref{Eq:Rotating_2comp_BEC}
and introducing an additional energy penalty associated with vortex cores.

\section{Monte-Carlo simulations}
\label{sec:MC_simulations}

\subsection{Considerations for a spherical surface}
\label{sec:spherical}

Computer simulations of Coulomb interactions are generally difficult to
perform due to the long-ranged nature of the interaction. Several
techniques have been presented to deal with the complications that
arise~\cite{Perram_PhysicaA_1981, Greengard_JComputPhys_1987, Maggs_PRL_2002}.
We have performed large-scale Monte-Carlo simulations of the system
described in Eqs.~\eqref{Eq:Model_Z} and \eqref{Eq:Model_S} on a
spherical surface. For other simulations on a spherical surface, see
Refs.~\onlinecite{Caillol_JStatPhys_1982, Caillol_PRB_1986,
  Caillol_JChemPhys_1991, O'Neill_PRB_1993, Dodgson_PRB_1997,
  Perez-Garrido_PRB_1998, Moore_PRL_1999}. This may seem like a
brute-force approach since the workload of the simulations scales as
$\mathcal{O}(N^2)$.  However, the benefit is that there are no
boundaries, the implementation is relatively easy, and there is no need
to constrain the particles to move on a lattice. However, one must also
be aware that simulation results may differ due to effects induced by
topology. For instance, the triangular crystalline ground state of a
2D one-component plasma will necessarily include a number of dislocations
and disclinations on a sphere. These defects are
not present in the ground state when the one-component plasma is located
on the plane~\cite{Dodgson_PRB_1997, Perez-Garrido_PRB_1997}.

We consider a sphere with radius $R$, with origin defined as the center of the
sphere such that all particle position vectors $\bfw_a$ and $\bfz_i$ are
radial vectors with fixed magnitude $R$ in three dimensions. The
distance between the particles is measured along the
chord~\cite{Caillol_PRB_1986, Caillol_JChemPhys_1991}
\begin{equation}
|\bfr_i-\bfr_j| = 2R \sin\left(\frac{\psi_{ij}}{2}\right),
\label{Eq:Chord}
\end{equation}
where
\begin{equation}
\psi_{ij} = \arccos(\hat{\bfr}_i\cdot\hat{\bfr}_j)
\label{Eq:Chord_angle}
\end{equation}
is the chord angle between the two particles at $\bfr_i$ and $\bfr_j$
with unit vectors $\hat{\bfr}_i$ and $\hat{\bfr}_j$, respectively. We
may now rewrite the model in Eq.~\eqref{Eq:Model_S}
on the surface of a unit sphere as
\begin{align}
V = \frac{1}{2}\Bigg [& Q_2^2 \sum_{a<b=1}^N \tilde{v}_{ww}(\hat{\bfw}_a\cdot
  \hat{\bfw}_b) + Q_2^2 \sum_{a,i=1}^N \tilde{v}_{zw}(\hat{\bfz}_i\cdot\hat{\bfw}_a)\nonumber\\
&+ (Q_1^2+Q_2^2)\sum_{i<j=1}^N \tilde{v}_{zz}(\hat{\bfz}_i\cdot\hat{\bfz}_j)\Bigg],
\label{Eq:Model_S_sphere}
\end{align}
with interactions given by
\begin{align}
\tilde{v}_{zz}(\hat{\bfr}_i\cdot\hat{\bfr}_j) =& \tilde{v}_{ww}(\hat{\bfr}_i\cdot\hat{\bfr}_j) \notag \\
=& \left\{ \begin{array}{rr}
\infty, & \psi_{ij} \leq d/R, \\
-\ln(1-\hat{\bfr}_i\cdot\hat{\bfr}_j), & \psi_{ij} > d/R, \\
\end{array} \right.
\label{Eq:Same_component_interaction_sphere}
\end{align}
and
\begin{equation}
\tilde{v}_{zw}(\hat{\bfr}_i\cdot\hat{\bfr}_j) = \left\{ \begin{array}{rr}
\infty, & \psi_{ij} \leq d/R, \\
\ln(1-\hat{\bfr}_i\cdot\hat{\bfr}_j), & \psi_{ij} > d/R. \\
\end{array} \right.
\label{Eq:Different_component_interaction_sphere}
\end{equation}
Note that the interaction $V_{z, \textnormal{BG}}$ in Eq.~\eqref{Eq:Model_S} between the neutralizing background and the
excess charge of type 1 becomes a constant term on the sphere, so we disregard it in Eq.~\eqref{Eq:Model_S_sphere}.

The dimensionless density of particles on the sphere is given by the packing fraction
$\eta = Ns/A$ where $s = A \sin^2(d/4R)$ is the area of a hard disk of
diameter $d$ on the sphere of area $A = 4\pi R^2$. In the simulation,
we use a unit sphere with $R = 1$.

As explained in Appendix~\ref{App:General_dielectric_constant}, in
order to account for screening
properties when particles interact by two interactions simultaneously, we
measure a general inverse dielectric constant, $\epsilon_{(a_1,
  a_2)}^{-1}$, given by
\begin{equation}
\epsilon_{(a_1, a_2)}^{-1} = a_1^2 \epsilon_{11}^{-1} + 2a_1a_2
\epsilon_{12}^{-1} + a_2^2 \epsilon_{22}^{-1},
\label{Eq:General_inverse_dielectric_constant_two_channels}
\end{equation}
where
\begin{equation}
\epsilon_{\mu \nu}^{-1} = \delta_{\mu \nu} - \frac{\pi}{A} \left \langle \bfM_\mu\cdot\bfM_\nu\right \rangle,
\label{Eq:Inverse_dielectric_constant_1_two_channels}
\end{equation}
is a type specific inverse dielectric constant, $a_1$ and
$a_2$ are type-dependent weights for the contributions of the different $\epsilon^{-1}_{\mu \nu}$ (which are determined by the values of both types of charge carried by the test particles for which screening is being measured), and where $\bfM_1$ and $\bfM_2$ are the dipole moments
for charges of type 1 and type 2, respectively, given by
\begin{equation}
\bfM_1 = Q_1 R \sum_{i=1}^{N}\hat{\bfz}_{i},
\label{Eq:Dipole_moment_type_1}
\end{equation}
\begin{equation}
\bfM_2 = Q_2 R\left(\sum_{i=1}^{N}\hat{\bfz}_{i} - \sum_{a=1}^{N}\hat{\bfw}_{a}\right).
\label{Eq:Dipole_moment_type_2}
\end{equation}
Note that the type 2 inverse dielectric constant, $\epsilon_{22}^{-1}$, is the same dielectric constant as was used
when studying the two-component neutral Coulomb plasma on a spherical
surface~\cite{Caillol_PRB_1986, Caillol_JChemPhys_1991}. In addition
to measuring the screening properties, the inverse dielectric constant may be used to identify the existence of a BKT-transition if
it exhibits a universal discontinuous jump at the critical point,
according to Eq.~\eqref{Eq:Universal_jump}.

In addition to the inverse dielectric constant, we also measure the fourth-order modulus, $\gamma$~\cite{Minnhagen_PRB_2003, Borkje_PRB_2005}.
This quantity may be used to verify a discontinuous jump in the inverse dielectric constant without making any
\textit{a priori} assumptions regarding the character of the phase
transition. As explained in detail in
Appendix~\ref{App:Fourth_order_derivative}, a negative $\gamma$ at the
phase transition in the thermodynamic limit implies that the inverse
dielectric constant jumps to zero discontinuously. As for the
inverse dielectric constant, we use a general fourth-order modulus to
account for the two interactions,
\begin{equation}
\gamma_{(a_1, a_2)} = \sum_{\mu,\nu,\rho,\sigma = 1}^2a_{\mu}a_{\nu}a_{\rho}a_{\sigma}\gamma_{\mu\nu\rho\sigma},
\label{Eq:General_fourth_order_modulus_two_interactions}
\end{equation}
where
\begin{eqnarray}
\gamma_{\mu\nu\rho\sigma} &=& \left(\frac{\pi}{R^2}\right)^2 \left[ \left\langle
\bfM_{\mu}\bfM_{\nu}\right\rangle\left\langle  \bfM_{\rho}\bfM_{\sigma}\right\rangle \right. \notag \\
&& \qquad \qquad \left. -3\left\langle M_{\mu,z}M_{\nu,z}M_{\rho,z}M_{\sigma,z}\right\rangle \right]
.
\label{Eq:Specific_fourth_order_modulus_two_interactions}
\end{eqnarray}
The explicit derivation of
Eqs.~\eqref{Eq:General_fourth_order_modulus_two_interactions} and
\eqref{Eq:Specific_fourth_order_modulus_two_interactions}, is given in Appendix~\ref{App:Fourth_order_derivative}.

\subsection{Details of the Monte-Carlo simulations}
\label{sec:MC_details}

The Monte-Carlo updating scheme consists of trial moves for one or two
particles at the same time, to a randomly
chosen new location on the surface of the sphere. The change in the action
Eq.~\eqref{Eq:Model_S_sphere} was calculated and the move was accepted or rejected according to
the Metropolis-Hastings algorithm~\cite{Metropolis_JChemPhys_1953, Hastings_Biometrika_1970}.
The trial moves were performed in three different ways. The first way
was to move a single particle to a new random
location uniformly over the total surface. The second way was to move a single particle to a new random
location uniformly within some short distance, adjusted to
yield a high acceptance rate. The last trial move was to move a
nearest-neighbor pair of one $z$-particle and one $w$-particle together, to a
random new location uniformly within some short distance, adjusted to
yield a high acceptance rate, and with a random new orientation. In order to
straightforwardly ensure detailed balance, we additionally required the two
particles to mutually be nearest-neighbors both in the old and the
new configuration. And to ensure ergodicity, the pair-move must be
mixed with a number of single-particle moves. All of these moves were found to be
essential in order to have fast thermalization as well as short
autocorrelation times for the cases considered here. Pseudorandom numbers were generated by the Mersenne-Twister algorithm~\cite{Matsumoto_ACMTMCS_1998} and the sampled data were postprocessed using Ferrenberg-Swendsen reweighting
techniques~\cite{Ferrenberg_PRL_1988, Ferrenberg_PRL_1989}.

\subsection{Results}
\label{sec:results}

Motivated by its relevance to the fractional quantum Hall effect (in particular, the $\nu=1/2$ MR state), we focus on analyzing the screening properties of this system at $Q_1 = 2$ ($M=2$)~\cite{Bonderson_PRB_2011}. We also perform simulations in the neutral
two-component Coulomb gas case at $Q_1 = 0$ ($M=0$) in order to provide a check on the numerics, as well as for comparison with
the $Q_1 = 2$ case. Furthermore, the system is also studied for a
number of values of the packing fraction, $\eta$ to extract the
screening properties in the low-density limit.

For the two cases of $Q_1$ and the values of $Q_2$ studied below, the quantities
$\epsilon_{11}^{-1}$ and $\epsilon_{12}^{-1}$ were found to be zero,
within statistical uncertainty, and except for a small finite-size effect
when system size $N$, was small. Thus, we focus on the results for
$\epsilon_{22}^{-1}$ as this was the only term in
Eq.~\eqref{Eq:General_inverse_dielectric_constant_two_channels} that
contributed to the general inverse dielectric constant,
$\epsilon_{(a_1, a_2)}^{-1}$. This means that screening properties of
particles that interact with charges of both types, are determined by
the charges of type 2, only. Note also that when $\epsilon_{11}^{-1} =
0$, the unconventional Coulomb plasma will screen test particles with
charge of type 1, only.

\begin{figure}[tbp]%%[htbp]
\includegraphics[width=\columnwidth]{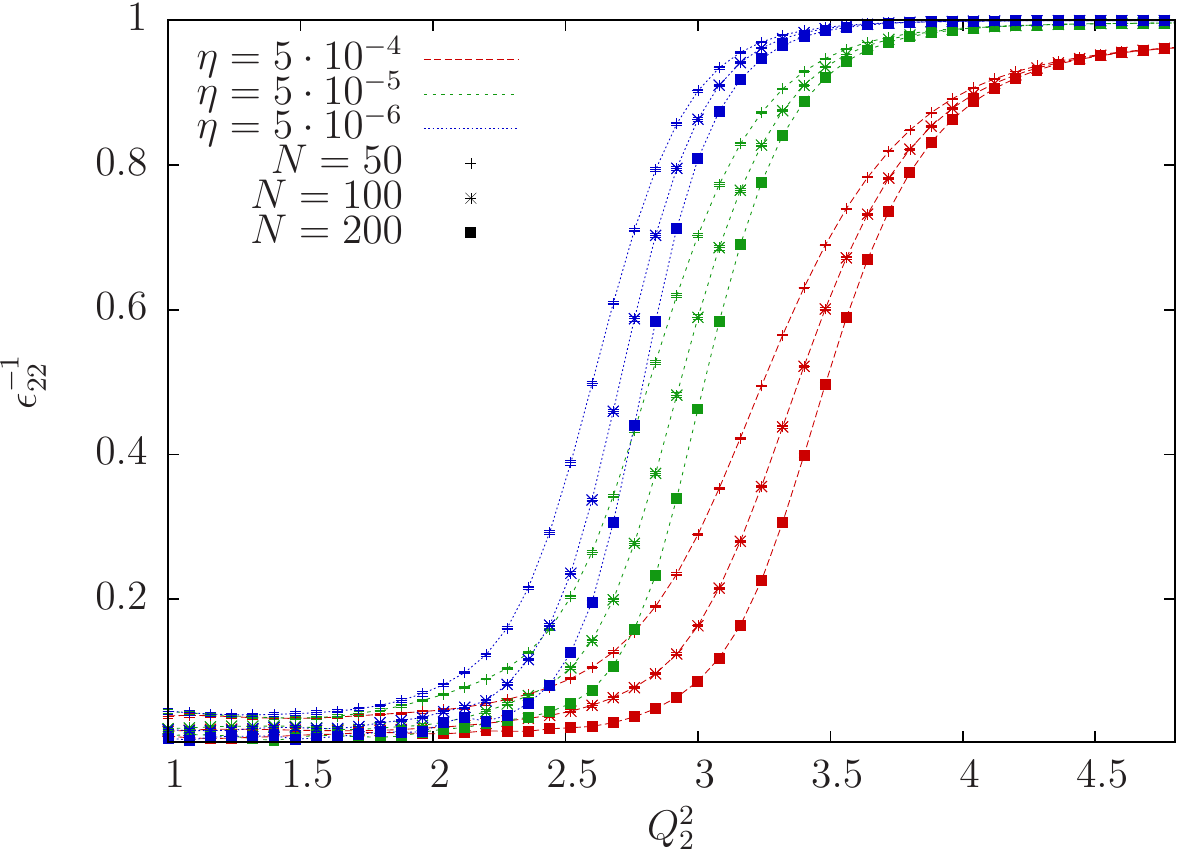}
\caption{(Color online) Plot of the inverse dielectric constant
  $\epsilon_{22}^{-1}$ for the model in Eq.~\eqref{Eq:Model_S} with $Q_1 = 0$
  and $1 \leq Q_2^2 \leq 4.8$. Results are presented for three different
  values of packing fraction $\eta$ and three different values of
  system size $N$.}
\label{Fig:Effect_of_eta_and_N}
\end{figure}

In Fig.~\ref{Fig:Effect_of_eta_and_N},
we plot $\epsilon_{22}^{-1}$ in
the relevant range of $Q_2^2$ when the two-component neutral Coulomb gas ($Q_1 = 0$)
is known to have a BKT transition.
At small values of $Q_2^2$, the system is in the screening phase
where $\epsilon_{22}^{-1} \approx 0$.
The reason for the $\approx$ sign rather than an equal sign
is that there is a mainly size-dependent offset from $\epsilon_{22}^{-1} = 0$, because perfect screening is not possible
with a small number of charges.
For large $Q_2^2$ there is a phase in which charges of different components form
tightly bound dipoles and the Coulomb gas turns into an insulator
where $\epsilon_{22}^{-1} \approx 1$.  Here, there is a mainly density-dependent offset
from $\epsilon_{22}^{-1} = 1$ because the polarizability
of the system increases with density, since the hard-core diameter
$d$ yields a minimum distance between the charges
in the dipoles. The plot in Fig.~\ref{Fig:Effect_of_eta_and_N}
indeed shows that the charge-unbinding transition
is dependent on the number of particles in the system, as well as the size of the hard disk charges. When $N$ increases,
the onset of a finite value in $\epsilon_{22}^{-1}$ moves to higher values of $Q_2^2$. However, when we reduce $\eta$, the value
of $Q_2^2$ at onset of $\epsilon_{22}^{-1}$ becomes smaller. Thus, this figure illustrates that understanding the behavior in
both limits $N \rightarrow \infty$ as well as $\eta \rightarrow 0$ is not straightforward.

\begin{figure}[tbp]%%[htbp]
\includegraphics[width=\columnwidth]{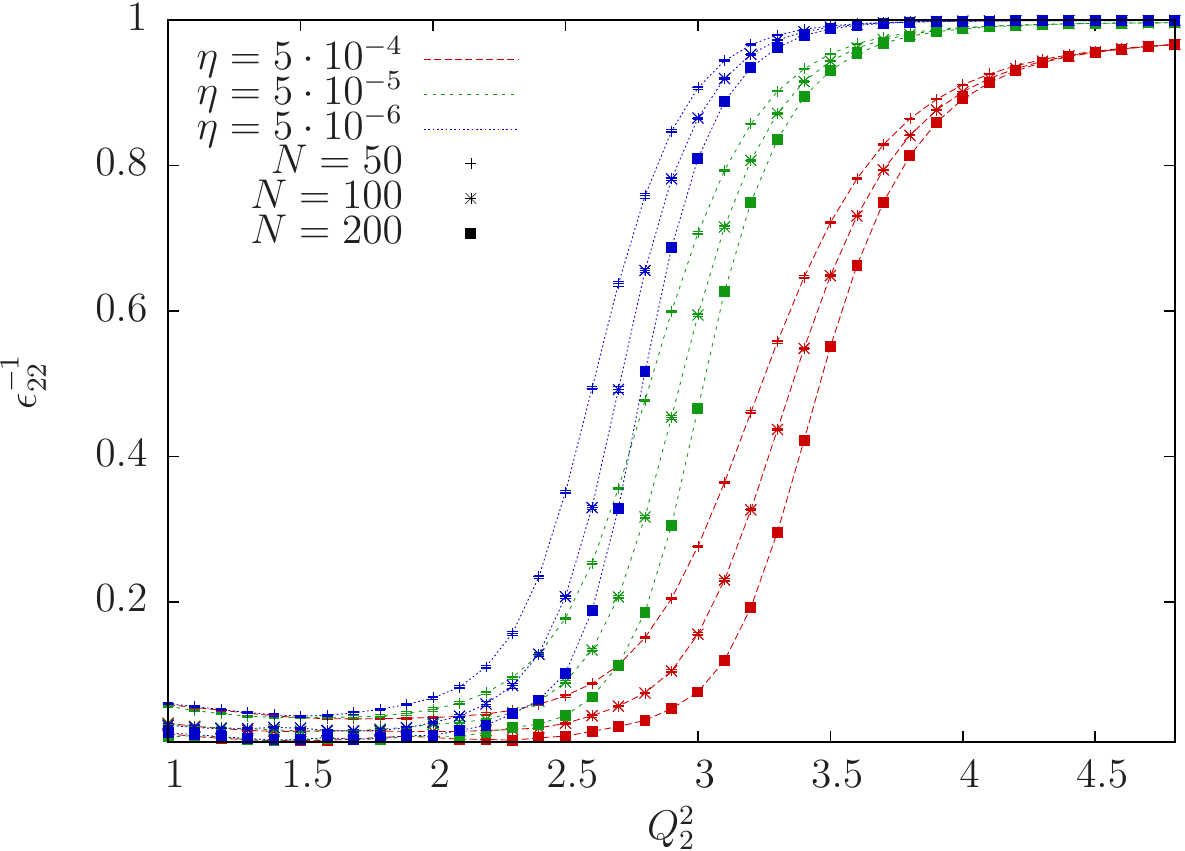}
\caption{(Color online) Plot of the inverse dielectric constant
  $\epsilon_{22}^{-1}$ for the model in Eq.~\eqref{Eq:Model_S} with $Q_1 = 2$
  and $1 \leq Q_2^2 \leq 4.8$. Results are presented for three different
  values of packing fraction $\eta$ and three different values of
  system size $N$.}
\label{Fig:Effect_of_eta_and_N-M_2}
\end{figure}

In Fig.~\ref{Fig:Effect_of_eta_and_N-M_2}, results for the same case as in Fig.~\ref{Fig:Effect_of_eta_and_N} are
presented, but with $Q_1 = 2$. The results for $Q_1 = 0$ and $Q_1 = 2$ are very similar, both qualitatively and quantitatively.
Thus, the screening properties with respect to charge of type 2 of the unconventional Coulomb plasma when $Q_1 = 2$ are very similar to
the well-studied two-component neutral Coulomb gas.

\begin{figure}[tbp]%%[htbp]
\includegraphics[width=0.768\columnwidth]{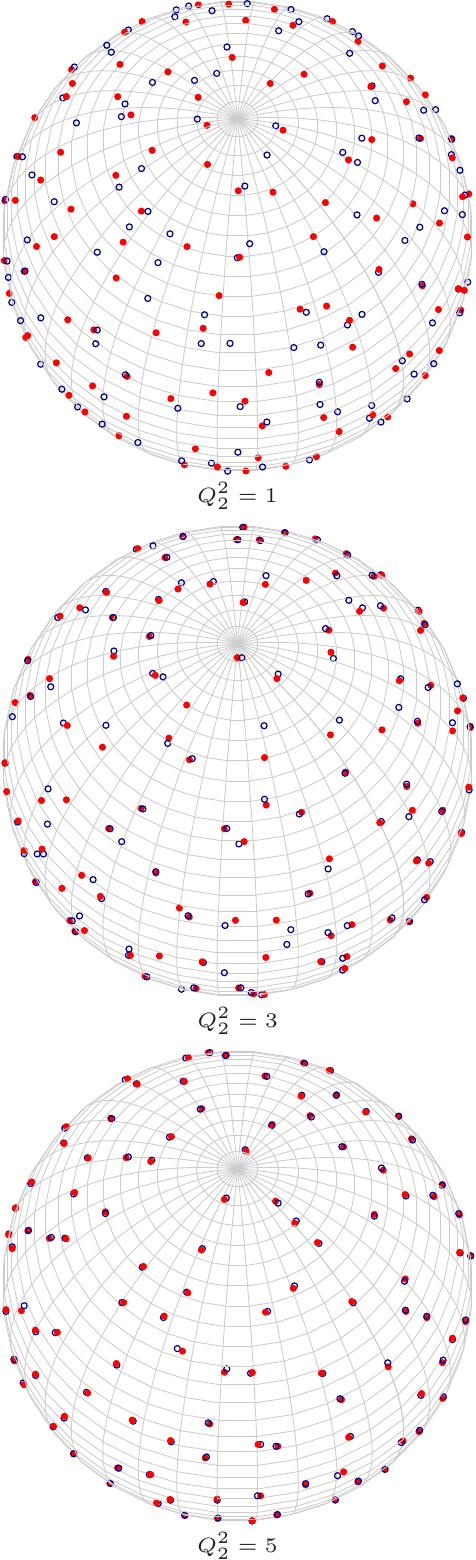}
\caption{(Color online) Snapshots of the charge configuration at $Q_2^2 = 1, 3, 5$ when $Q_1 = 2$, $\eta = 5\cdot 10^{-4}$
and $N = 200$. Red markers represent $w$-particles, while blue markers represent $z$-particles. The marker diameters are
about $5$ times larger than hard disk diameter $d$.}
\label{Fig:Charge_grid}
\end{figure}

To get a qualitative picture of the type 2 charge binding of
the unconventional plasma, three snapshots of the charge configuration
when $Q_1 = 2$, $\eta = 5\cdot 10^{-4}$, and $N = 200$ is given in
Fig.~\ref{Fig:Charge_grid}. When $Q_2^2 = 1$, deep into the screening
phase of the system (see Fig.~\ref{Fig:Effect_of_eta_and_N-M_2}), most
charges are free and only a small fraction of the charges may be said
to form closely bound dipoles. At $Q_2^2 = 3$, which is the relevant value for the Ising-type quantum Hall states, the system is closer to the unbinding transition and a larger fraction (though not all) of the particles are bound in
dipoles. At $Q_2^2 = 5$, deep in the type 2 insulating region, all particles form closely
bound dipoles and the ability to screen type 2 test charges is lost.

Although it is clear from
Figs.~\ref{Fig:Effect_of_eta_and_N},\ref{Fig:Effect_of_eta_and_N-M_2}
that there is a transition between a screening phase and an insulating phase,
it is not easy to spot the transition point in the curves in these figures,
which look rather smooth. Therefore, we must make some assumptions
about the nature of the transition in order to identify it.

For the case $Q_1 = 0$, where the transition is known to be a BKT transition,
it is natural to follow a method that was proposed in Ref.~\onlinecite{Weber_PRB_1988}.
At the BKT critical point, $\epsilon_{22}^{-1}$ scales
logarithmically with $N$ for large $N$. It taked the following
finite-size scaling form:
\begin{equation}
\epsilon_{22}^{-1}(N) = \epsilon_{22}^{-1} (\infty) \left[1 + \frac{1}{\ln(N)+C} \right],
\label{Eq:Finite_size_scaling}
\end{equation}
where $\epsilon_{22}^{-1} (\infty)$ is the value of $\epsilon_{22}^{-1}(N)$ when $N \rightarrow \infty$ and $C$ is an undetermined
constant. Least-squares curve-fitting to Eq.~\eqref{Eq:Finite_size_scaling} may be performed for various sizes $N$ with $C$ and
$\epsilon_{22}^{-1} (\infty)$ as free parameters at fixed values of $Q_2^2$. The critical point is then estimated as the value
of $Q_2^2$ which exhibits the best fit to
Eq.~\eqref{Eq:Finite_size_scaling}. Additionally, for a
BKT-transition, the value of $\epsilon_{22}^{-1} (\infty)$ obtained at the best
fit, must correspond with the universal jump condition, $Q_{2,\text{c}}^2 \epsilon_{22}^{-1} (\infty) = 4$,
cf. Eq.~\eqref{Eq:Universal_jump}. Details of this procedure are
given in Appendix~\ref{App:Weber_Minnhagen}.

For $Q_1 = 2$, motivated by the similarity between
Figs.~\ref{Fig:Effect_of_eta_and_N}, \ref{Fig:Effect_of_eta_and_N-M_2},
we assume that the transition is also a BKT transition.
We again look for the $Q_2^2$ value at which the system best
fits Eq. \eqref{Eq:Finite_size_scaling}. Since we are able to find a value
at which there is a very good fit to this form, we conclude that our
assumption was justified.

\begin{figure}[tbp]%%[htbp]
\includegraphics[width=\columnwidth]{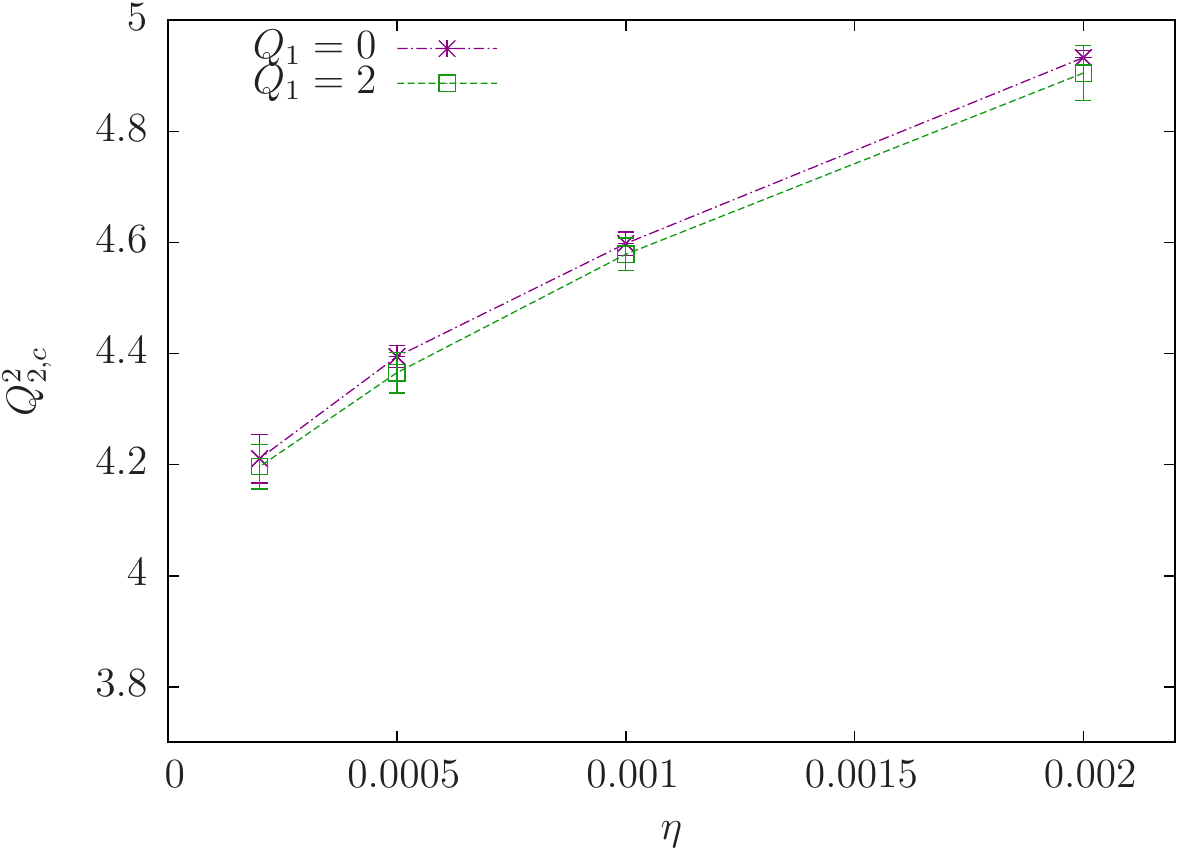}
\caption{(Color online) The critical value of $Q_2^2$ found by curve fitting to Eq.~\eqref{Eq:Finite_size_scaling} with
two free parameters. Results are presented for four values of the packing fraction $\eta$ and for $Q_1 = 0$ and $Q_1 = 2$.
Fourteen system sizes in the range $20 \leq N \leq 2000$ have been used.}
\label{Fig:Critical_QQ_Weber_Minnhagen_method}
\end{figure}

\begin{figure}[tbp]%%[htbp]
\includegraphics[width=\columnwidth]{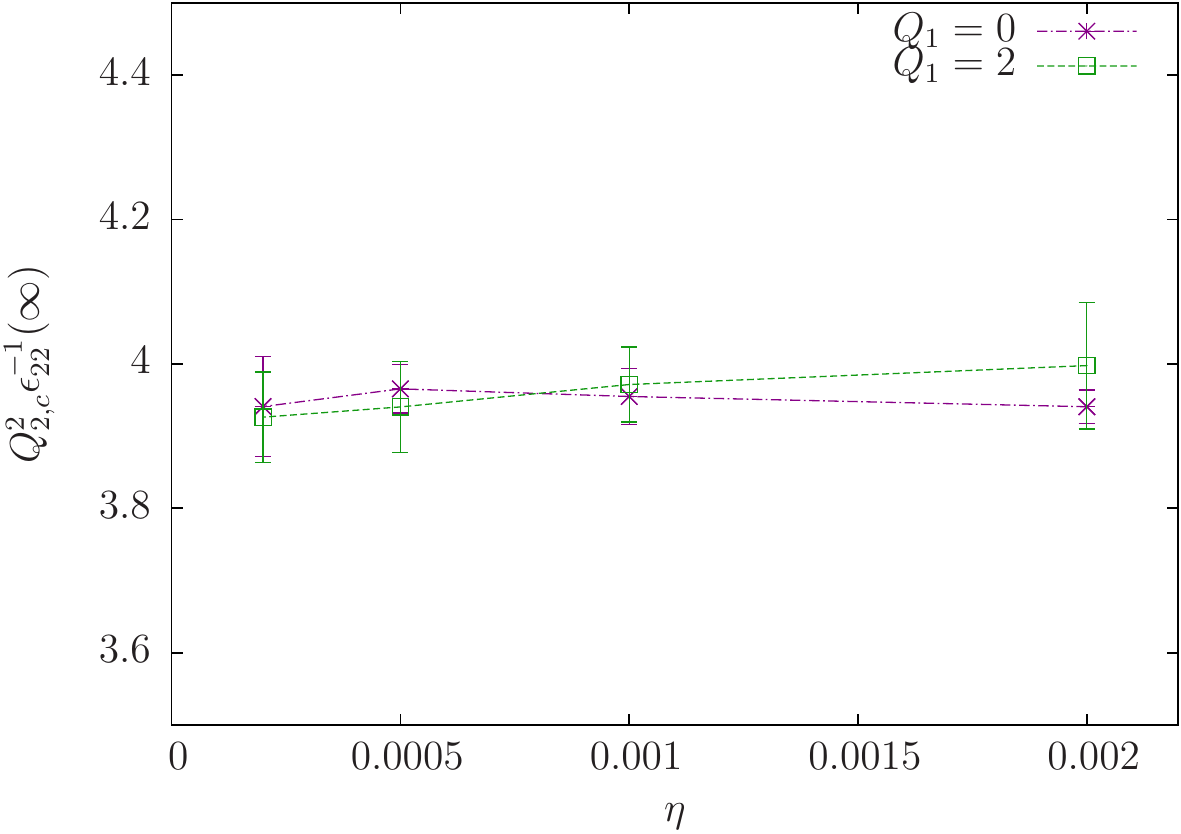}
\caption{(Color online) The universal jump value determined by curve fitting to Eq.~\eqref{Eq:Finite_size_scaling}
with two free parameters. Results are presented for four values of the packing fraction $\eta$ and for $Q_1 = 0$ and
$Q_1 = 2$. Fourteen system sizes in the range $20 \leq N \leq 2000$ have been used.}
\label{Fig:Universal_jump_Weber_Minnhagen_method}
\end{figure}

In Fig.~\ref{Fig:Critical_QQ_Weber_Minnhagen_method}, we present results for the critical coupling $Q_{2,\text{c}}^2$ for four
different densities $\eta=0.0002,0.0005,0.001,0.002$, for $Q_1 = 0$ and $Q_1 = 2$. The results for $Q_1 = 0$ reproduce the
main features of the two-component Coulomb gas, namely that $Q_{2,\text{c}}^2 = 4$ when density is low and that $Q_{2,\text{c}}^2$
increases when density increases. These results also correspond well with earlier results in Refs.~\onlinecite{Caillol_PRB_1986}
and \onlinecite{Orkoulas_JChemPhys_1996}. When $Q_1 = 2$, we find that the behavior of the critical temperature is very similar
to the $Q_1 = 0$ case, within statistical uncertainty. In addition, in Fig.~\ref{Fig:Universal_jump_Weber_Minnhagen_method},
results for the corresponding value of the parameter $\epsilon_{22}^{-1} (\infty)$ at the critical point is presented. The values
for both $Q_1 = 0$ and $Q_1 = 2$ are close to the universal value of $Q_{2,\text{c}}^2 \epsilon_{22}^{-1} (\infty)= 4$ for the BKT-transition.
Since the results for $Q_1=0$ (the standard Coulomb-plasma BKT-transition case) and $Q_1 = 2$ are essentially the same, we suggest
that the charge-unbinding transition for the unconventional Coulomb plasma indeed is a BKT-transition in the sense that the type 2 inverse dielectric
constant $\epsilon_{22}^{-1}$ exhibits logarithmic finite-size scaling and a discontinuous jump with a universal value, as
predicted by the BKT renormalization equations.

\begin{figure}[tbp]%%[htbp]
\includegraphics[width=\columnwidth]{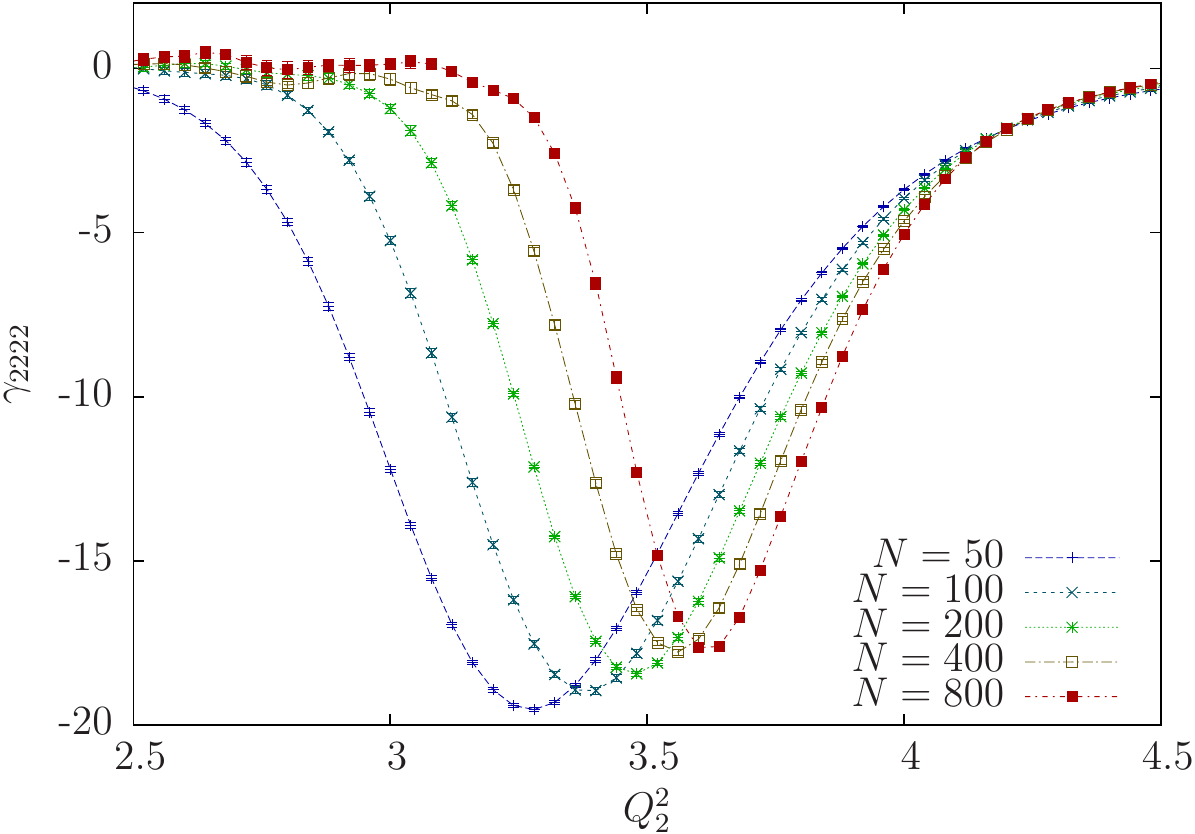}
\caption{(Color online) The fourth-order modulus $\gamma_{2222}$ as a
  function of coupling $Q_2^2$ for five different system sizes $N$, when $Q_1 = 2$ and
$\eta = 5\cdot 10^{-4}$.}
\label{Fig:Typical_bahavior}
\end{figure}

\begin{figure}[tbp]%%[htbp]
\includegraphics[width=\columnwidth]{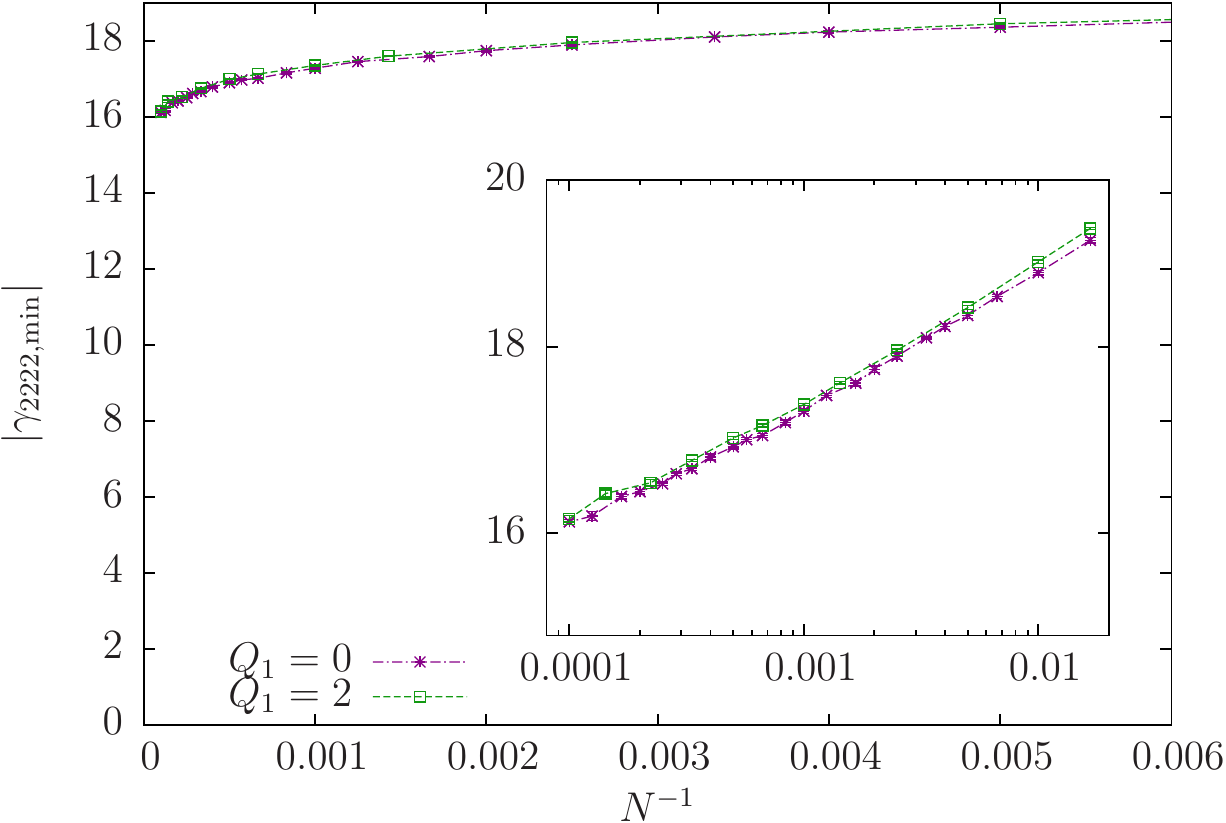}
\caption{(Color online) The size of the dip in the
  fourth-order modulus $|\gamma_{2222,
  \textnormal{min}}|$ as a function of inverse system size
$N^{-1}$. The packing fraction is $\eta = 5\cdot 10^{-4}$, and results
for $Q_1 = 0$ and $Q_1 = 2$ are shown. The inset shows the results on a
log-log scale. System sizes in the range $60 \leq
N \leq 10000$ are used.}
\label{Fig:Dip_fourth_order_modulus}
\end{figure}

\begin{figure}[tbp]%%[htbp]
\includegraphics[width=\columnwidth]{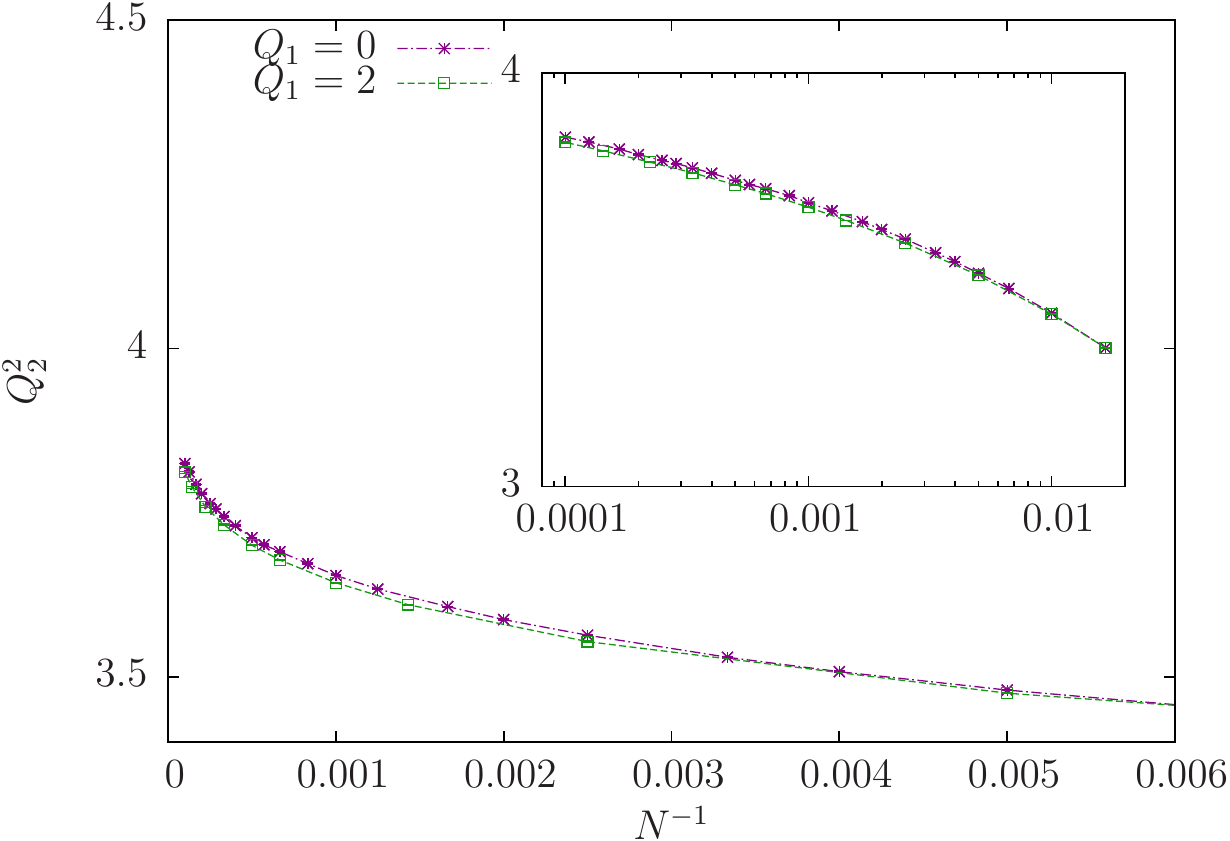}
\caption{(Color online) The coupling value at the minimum of the dip in the
  fourth-order modulus as a function of inverse system size
$N^{-1}$. The packing fraction is $\eta = 5\cdot 10^{-4}$, and results
for $Q_1 = 0$ and $Q_1 = 2$ are shown. The inset shows the results on a
log-log scale. System sizes in the size $60 \leq
N \leq 10000$ are used.}
\label{Fig:Dip_critical_qq}
\end{figure}

As an additional verification of the discontinuous jump in the
BKT-transition, we also study the fourth-order modulus $\gamma_{(a_1, a_2)}$,
presented in
Eqs.~\eqref{Eq:General_fourth_order_modulus_two_interactions} and
\eqref{Eq:Specific_fourth_order_modulus_two_interactions}. As for the
general inverse dielectric constant, we found that the only contributing term in the
sum of Eq.~\eqref{Eq:General_fourth_order_modulus_two_interactions} is
the term with all indices equal to 2, $\gamma_{2222}$. Illustrating the typical behavior of this quantity,
results for $\gamma_{2222}$ for a number of sizes when $\eta = 5\cdot 10^{-4}$ and $Q_1 = 2$ are presented in Fig.~\ref{Fig:Typical_bahavior}. Typically, $\gamma_{2222}$ exhibits a dip at a value of the coupling that can be associated
with the transition. As explained in Appendix~\ref{App:Fourth_order_derivative}, a
negative and finite dip in the limit when $N \rightarrow \infty$
signals the discontinuous jump in $\epsilon_{22}^{-1}$ that is a
characteristic feature of a BKT-transition. To this end, the size of
the dip in $\gamma_{2222}$ is plotted as a function of inverse system
size $N^{-1}$ in Fig.~\ref{Fig:Dip_fourth_order_modulus} in the case
when $\eta = 5\cdot 10^{-4}$. The size of the dip $|\gamma_{2222,
  \textnormal{min}}|$ decreases when $N$ increases towards the
thermodynamic limit. However, assuming power-law dependence of $|\gamma_{2222,
  \textnormal{min}}|$, the positive curvature in the log-log plot
indicates a nonzero value of $|\gamma_{2222,
  \textnormal{min}}|$ when $N
\rightarrow \infty$, verifying a discontinuous jump in
$\epsilon_{22}^{-1}$, as expected for a BKT-transition. Again, we find that
the results for $Q_1 = 2$ are very similar to $Q_1 = 0$.
We also associate the coupling value of the minimum in the dip in $\gamma_{2222}$ with
the critical point and the results are shown in Fig.~\ref{Fig:Dip_critical_qq} in the case when $\eta = 5\cdot 10^{-4}$. Clearly, the position
of the dip moves towards higher values of $Q_2^2$ when the system size increases. However,
the evolution towards $N^{-1} = 0$ is too slow to make a sharp determination of $Q_2^2$ in this
limit as also noted before~\cite{Minnhagen_PRB_2003, Borkje_PRB_2005}. With this method, we
are not able to verify that $Q_{2,\text{c}}^2 \approx 4.4$, as was found above in Fig.~\ref{Fig:Critical_QQ_Weber_Minnhagen_method} for this density.

\begin{figure}[tbp]%%[htbp]
\includegraphics[width=\columnwidth]{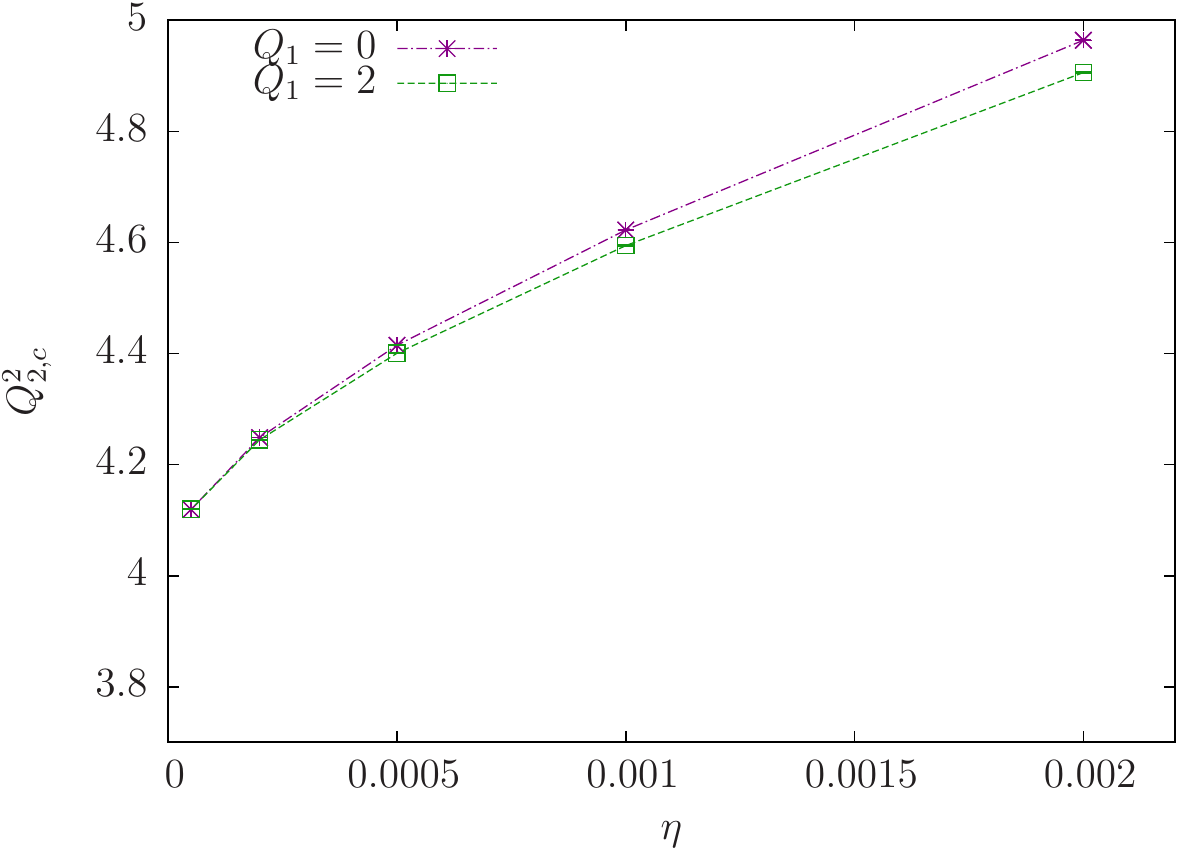}
\caption{(Color online) The critical value of $Q_2^2$ found by curve-fitting to Eq.~\eqref{Eq:Finite_size_scaling} with one free
  parameter. Results are presented for five values of the packing
  fraction $\eta$ and for two values of $Q_1$.}
\label{Fig:Critical_QQ_ofp_Weber_Minnhagen_method}
\end{figure}

By assuming a universal value of the discontinuous jump for a BKT-transition, we
may determine the critical point of the BKT-transition using
Eq.~\eqref{Eq:Finite_size_scaling} with only one free parameter as
described in Appendix~\ref{App:Weber_Minnhagen}. The results are given in
Fig.~\ref{Fig:Critical_QQ_ofp_Weber_Minnhagen_method}. The critical
values of $Q_2^2$ are very similar to what was obtained in
Fig.~\ref{Fig:Critical_QQ_Weber_Minnhagen_method}, but are
determined with greater accuracy. For both cases, the critical point
appears at higher $Q_2^2$ when density increases. However, $Q_{2,\text{c}}^2$ is
systematically lower at $Q_1 = 2$ compared to $Q_1 = 0$.

For the range of small densities that we have investigated, the Monte-Carlo results for
the unconventional Coulomb plasma with $Q_1 = 2$ are rather conclusive. This plasma undergoes a charge-unbinding
transition that should be regarded as a BKT-transition in the sense
that the inverse dielectric constant of type 2 exhibits the well-established signatures of a BKT-transition. Specifically, there is a
density-dependent critical point $Q_{2,\text{c}}^2$ that separates a phase where
particles of different species form bound pairs at high values of $Q_2^2$
from a phase where particles of different species are free at low
values of $Q_2^2$. For test particles carrying type 2 charge, the
high-$Q_2^2$ phase is unscreened, whereas the low-$Q_2^2$ phase is screened.

\begin{figure}[tbp]%%[htbp]
\includegraphics[width=0.768\columnwidth]{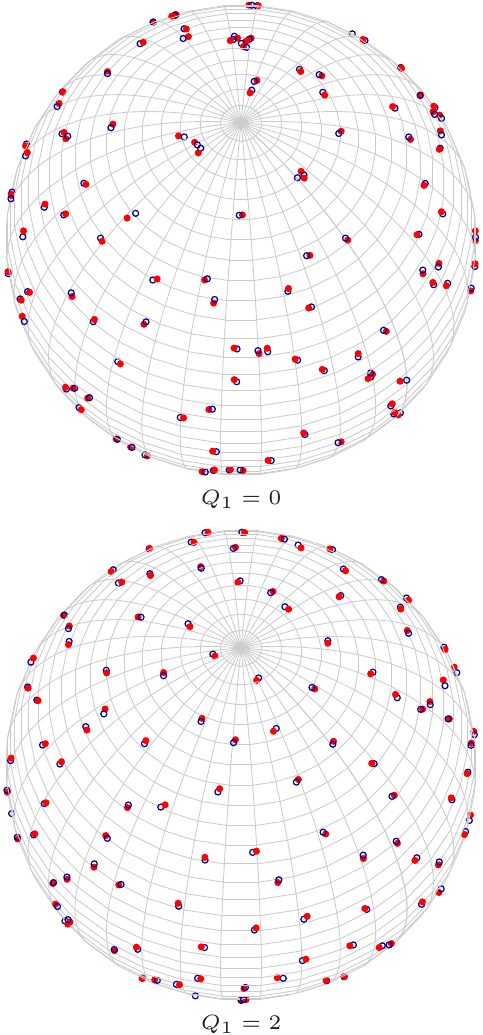}
\caption{(Color online) Snapshots of the charge configuration at $Q_1 = 0$ and $Q_1 = 2$
when $Q_2^2 = 7$, $\eta = 2\cdot 10^{-3}$, and $N = 200$. Red markers are $w$-particles and blue markers
are $z$-particles. The marker diameters are about 2.5 times larger than hard disk diameter $d$.}
\label{Fig:Charge_grid_bounded_phase}
\end{figure}

\begin{figure}[tbp]%%[htbp]
\includegraphics[width=\columnwidth]{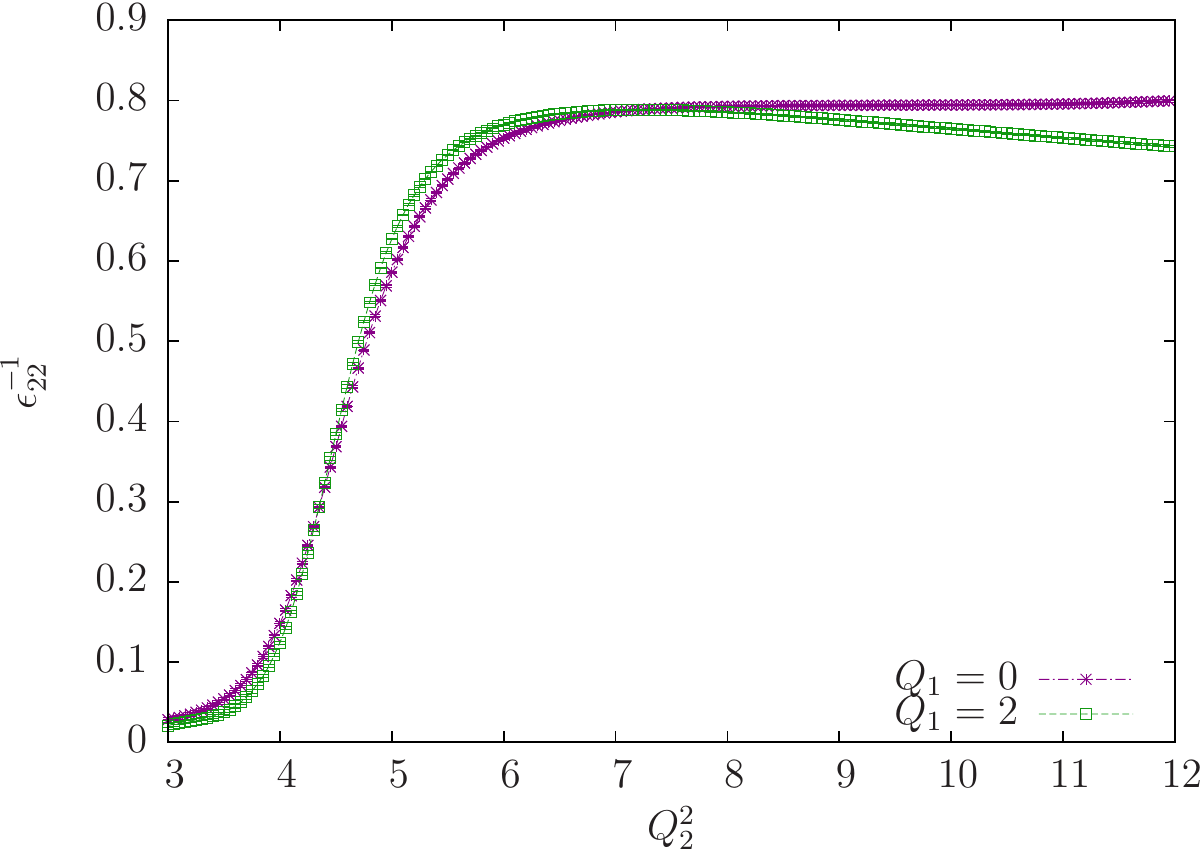}
\caption{(Color online) Plot of the type 2 inverse dielectric constant for $Q_1 = 0$ and $Q_1 = 2$ with $N = 100$,
$\eta = 5\cdot 10^{-3}$ in the range $3 \leq Q_2^2 \leq 12$.}
\label{Fig:Inverse_dielectric_constant_highdens}
\end{figure}

The results presented so far show that the behavior when $Q_1 = 0$ and $Q_1 = 2$ are quite similar.
However, in the phase with bounded dipoles, when charges of type 2 are
not screened, the cases $Q_1=0$ and $Q_1=2$ behave rather differently. We first
consider the case when $Q_1 = 0$. When charges are bound, this system consists of $N$
dipoles that interact by dipole-dipole interactions. Consequently,
these dipoles tend to form clusters with increased dipole strength, i.e. higher
values of the coupling or the density~\cite{Orkoulas_JChemPhys_1996, Lidmar_PRB_1997}.
In Fig.~\ref{Fig:Charge_grid_bounded_phase}, a snapshot of a $Q_1 = 0$ configuration
with $N = 200$, $Q_2^2 = 7$ and $\eta = 2\cdot 10^{-3}$ is shown, where some dipoles
are seen to form clusters. In the case when $Q_1 = 2$, the type 2 interactions are effectively reduced to dipole-dipole interactions,
similar to the $Q_1 = 0$ case. {\it However, the logarithmic
  interactions of type 1 charges remain}. Neglecting the weaker
dipole-dipole interactions among dipoles of type two, the dipoles now essentially form elementary constituents
with charge $Q_1$ interacting logarithmically. Effectively, the two-component unconventional plasma is reduced
to a one-component plasma where the particles carry charge of type 1 and a (neutral) dipole of type 2.
When $Q_1 =2$ this plasma is in the liquid state, i.e. the tightly bound dipoles do
not form an ordered state with a broken translational or orientational symmetry.
Also, the logarithmic interaction of type 1 charge will prevent the dipoles from
forming clusters. A snapshot of the state with bounded dipoles when $Q_1 = 2$ is
shown in Fig.~\ref{Fig:Charge_grid_bounded_phase} and the qualitative difference from
the case when $Q_1 = 0$ is clearly seen. Quantitatively, this is seen by the behavior of
$\epsilon_{22}^{-1}$, presented in Fig.~\ref{Fig:Inverse_dielectric_constant_highdens}. When
$Q_1 = 0$, dipole-dipole interactions at short distances will reduce the
fluctuations in the dipole moment resulting in a weakly increasing $\epsilon_{22}^{-1}$ inside
the bounded phase. On the other hand, when $Q_1 = 2$ the logarithmic interaction of type 1 charge
will keep the dipoles at some distance from each other, thus the fluctuations of a dipole are
not much restricted by the surrounding dipoles. Moreover, the strength of the dipoles increases
with $Q_2^2$ and a reduction in $\epsilon_{22}^{-1}$ follows. The qualitative difference between
the cases $Q_1 = 0$ and  $Q_1 = 2$ is an effect due to the minimum separation of charges at finite
density originating with the hard cores, and it will vanish in the limit $\eta \to 0$.

\section{Conclusions}
\label{sec:conclusions}

We have shown that the unconventional Coulomb plasma analyzed in this
paper, where particles can carry two distinct types of Coulombic
charge, will screen test particles with charges of both types for the case most relevant for the plasma analogy of Ising-type fractional quantum Hall
states, i.e. when there is one species of particles that carry type 1 charge $Q_1 = 2$ ($M=2$) and type 2 charge $Q_2 = \sqrt{3}$ and another species of particles that carry only type 2 charge $-Q_2$. For test particles carrying both types of charge,
screening will cease to occur at
$Q_2^2 = Q_{2,\text{c}}^2 \approx 4$ in the limit of small density, when $Q_1 = 2$.
For higher values of $Q_2^2$,
the system will continue to screen test particles that carry only type 1 charge,
but will not be able to screen test particles with type 2 charge.

One striking feature of these results is that $Q_{2,\text{c}}^2$ and the critical
behavior at this point hardly seem to depend on $Q_1$ when density is small.
This implies that the role of the type 1 interaction (which corresponds,
in quantum Hall wavefunction language, to the Laughlin-Jastrow factor
which accounts for the filling fraction of the system) is simply to
maintain the $z_i$-particles in a liquid state. Since its critical point is very far
away, the type 1 interaction leads to a weak, smooth dependence on $Q_1$.
The physics in the transition at $Q_{2,\text{c}}^2$ is then dominated by the
type 2 interaction. We therefore conjecture that our results hold for
all reasonable values of $M$ -- not only $M=0$ and $2$, the cases which
we have studied here, but also $M=1$ (which may be relevant to ultra-cold
trapped bosons) and larger values of $M$, possibly all the way up
to or near the critical value $M_{\text{c}} \approx 70$, below which the
one-component plasma of Eq.~(\ref{eqn:Laughlin-plasma}) is in the
metallic phase~\cite{de_Leeuw_PhysicaA_1982, Caillol_JStatPhys_1982, Choquard_PRL_1983, Franz_PRL_1994}.

\acknowledgments
We acknowledge useful discussions with K. B\o rkje, J.~S. H\o ye,
I.~B. Sperstad, B. Svistunov, and M. Wallin. E.~V.~H, E.~B., and
V.~G. thank Nordita for hospitality during the initial stage of this work.
E.~B., P.~B., and A.~S. thank the Aspen Center for Physics for hospitality and support under the NSF grant $\#1066293$.
E.~V.~H. thanks NTNU for financial support.
E.~B. was supported by Knut and Alice Wallenberg Foundation through the Royal Swedish Academy of Sciences Fellowship,
Swedish Research Council and by the National Science Foundation CAREER
Award No. DMR-0955902.
V.~G. was supported by NSF Grant No. PHY-0904017.
C.~N. was supported in part by the DARPA QuEST program.
A.~S. was supported by the Norwegian Research Council under Grant
No. 205591/V30 (FRINAT). The work was also supported through the
Norwegian consortium for high-performance computing (NOTUR).

\appendix

\section{Generalizing the inverse dielectric constant for multiple interactions}
\label{App:General_dielectric_constant}

In the unconventional plasma with two components that
interact with two different Coulomb-like interactions, we are free to insert test particles that may
interact with different charge strength through both interactions
simultaneously. Here, we will generalize the inverse dielectric
constant for such test particles. For consistency, we will also perform
the derivation on the surface of a sphere by expanding in spherical
harmonics. For a similar derivation, but with one interaction only and on a planar geometry, see
Refs.~\onlinecite{Olsson_PRB_1992, Olsson_PRB_1995}.

When an external test charge field is inserted in the system, the free energy in
the system will change according to the effective interaction among
the test charges,
\begin{equation}
\Delta F [\delta q] = \int \mathrm{d}\Omega \int
\mathrm{d}\Omega^{\prime} \sum_{\mu,\nu} \delta q_\mu (\theta, \phi)
U^{\textnormal{eff}}_{\mu \nu}(\hat{\bfr}\cdot\hat{\bfr}^{\prime}) \delta
q_\nu (\theta^{\prime},\phi^{\prime}).
\label{Eq:Free_energy_increase_by_effective_interaction}
\end{equation}
Here, the effective interaction between charges of type $\mu$ and $\nu$, is assumed to be of the form
$U^{\textnormal{eff}}_{\mu \nu} = U^{\textnormal{eff}}_{\mu \nu}(\hat{\bfr}\cdot\hat{\bfr}^{\prime})$, $\delta q_{\mu}
(\theta,\phi)$ is the test charge field for charges of type
$\mu$, and the integrations are over the solid angle $\mathrm{d}\Omega$. To correctly model the test particles as carrying charge of
different types, we write
\begin{equation}
\delta q_{\mu}(\theta,\phi) = a_{\mu} \, \delta q \, \rho(\theta,\phi),
\label{Eq:Test_charge_field}
\end{equation}
where $a_\mu$ is a type dependent factor that accounts for the
relative strength of charges of different types. For
instance, the choice $(a_1, a_2) = (Q_1/ M Q_2, 1) = (\sqrt{ 2 / 3M }, 1)$ describes
the test charges corresponding to quasiholes in the MR state, as given
in Eq.~(125) in Ref.~\onlinecite{Bonderson_PRB_2011}, which map to
particles in the plasma that carry charge $Q_1 / 2M = 1/\sqrt{2M}$ of
type 1 and charge $Q_2/2 = \sqrt{3}/2$ of type 2. Moreover, in Eq.~\eqref{Eq:Test_charge_field} $\delta q$ is a
common charge factor for all types such that $a_{\mu} \, \delta q$ is the
total charge of type $\mu$ carried by a test particle (which means
that $\delta q = \sqrt{3}/2$ in the example above), and $\rho(\theta,\phi)$
is the density field of the test particles.

It is now convenient to expand the interaction and density fields in
spherical harmonics. The test particle density field is expanded by
\begin{equation}
\rho(\theta, \phi) = \sum_{l=0}^{\infty}\sum_{m=-l}^{l}\rho_{l}^{m} Y_{l}^{m}(\theta, \phi),
\label{Eq:Spherical_harmonics_expansion}
\end{equation}
where
\begin{equation}
Y_{l}^{m}(\theta, \phi) =
\sqrt{\frac{(2l+1)(l-m)!}{4\pi(l+m)!}}P_{l}^{m}(\cos \theta) \, \e^{im\phi},
\label{Eq:Spherical_harmonics}
\end{equation}
and $P_{l}^{m}(x)$ are the associated Legendre polynomials. The
coefficients are given by
\begin{equation}
\rho_{l}^{m} = \int \mathrm{d}\Omega \, \rho(\theta, \phi)
Y_{l}^{m *}(\theta, \phi).
\label{Eq:Spherical_harmonics_coefficients}
\end{equation}
The effective interaction is expanded by using the addition theorem for spherical harmonics,
\begin{equation}
U^{\textnormal{eff}}_{\mu \nu}(\hat{\bfr}\cdot\hat{\bfr}^{\prime}) = \sum_{l=0}^{\infty}
\frac{4\pi}{2l+1} U^{\textnormal{eff}}_{\mu \nu, l} \sum_{m=-l}^{l}
Y_{l}^{m *}(\theta, \phi) Y_{l}^{m}(\theta^{\prime}, \phi^{\prime}).
\label{Eq:Effective_interaction_spherical_harmonics_expansion}
\end{equation}
Here, $U^{\textnormal{eff}}_{\mu \nu, l}$ are the Legendre coefficients of the interaction, given by
\begin{equation}
U^{\textnormal{eff}}_{\mu \nu, l} = \frac{2l+1}{2}\int_{0}^{\pi}
\mathrm{d} \theta \,\, \sin \theta \, U^{\textnormal{eff}}_{\mu \nu}(\cos \theta) P_l(\cos \theta),
\label{Eq:Legendre_coefficients}
\end{equation}
where $P_l(x)$ is the Legendre polynomial of order $l$. Now
Eq. \eqref{Eq:Free_energy_increase_by_effective_interaction} is written
\begin{equation}
\Delta F [\delta q] = \delta q^2 \sum_{l=0}^{\infty} \frac{4\pi}{2l+1} \sum_{\mu,\nu}
a_\mu U^{\textnormal{eff}}_{\mu \nu, l} a_\nu \sum_{m=-l}^{l} \rho_{l}^{m *} \rho_{l}^{m}.
\label{Eq:Free_energy_increase_by_effective_interaction_1}
\end{equation}
Hence, in the limit when the test charge field is infinitesimal,
$\delta q \rightarrow 0$, we find that
\begin{equation}
\frac{\partial^2 F [\delta q]}{\partial \delta q^2}\Bigg|_{\delta q =
  0} = \sum_{l=0}^{\infty} \frac{8\pi}{2l+1} \sum_{\mu, \nu}
a_\mu U^{\textnormal{eff}}_{\mu \nu, l} a_\nu \sum_{m=-l}^{l} \rho_{l}^{m *} \rho_{l}^{m}.
\label{Eq:Effective_second_derivative_free_energy}
\end{equation}

This derivative can also be calculated by inspection of the partition
function of the system perturbed with the external test charge
field. With $F[\delta q] = -\ln Z[\delta q]$ and a potential energy on
the form $V[\delta q] = V_0 + V_1[\delta q]$ where $V_0$ is the
potential energy of the unperturbed system and $V_1[\delta q]$ is the
contribution due to the test charge field, we find that
\begin{align}
\frac{\partial^2 F [\delta q]}{\partial \delta q^{2}}\Bigg|_{\delta
q = 0} =& \left \langle \frac{\partial^2 V_1 [\delta q]}{\partial \delta q^{2}}\Bigg|_{\delta
q = 0} \right \rangle \nonumber \\
&- \left \langle \left(\frac{\partial V_1 [\delta q]}{\partial \delta q}\Bigg|_{\delta
q = 0} \right)^2 \right \rangle.
\label{Eq:Bare_second_derivative_free_energy}
\end{align}
Here, we have also used that $\partial F [\delta q]/\partial \delta q|_{\delta
q = 0} = 0$, and the brackets denote statistical average with respect
to the unperturbed system. The test charges $\delta q_\mu(\theta, \phi)$ will interact with each
other as well as with the charge field $q_\mu(\theta, \phi)$. As for
the test charge field, the charge field is expanded according to
Eq.~\eqref{Eq:Spherical_harmonics_expansion} to yield
\begin{align}
V_1 [\delta q] =& \int \mathrm{d}\Omega \int
\mathrm{d}\Omega^{\prime} \sum_{\mu} \left[q_\mu (\theta, \phi) + \delta
  q_\mu (\theta, \phi)\right]\nonumber \\
& \times U(\hat{\bfr}\cdot\hat{\bfr}^{\prime}) \delta q_\mu (\theta^{\prime},\phi^{\prime})\nonumber\\
=& \sum_{l=0}^{\infty} \frac{4\pi}{2l+1} U_{l}
\sum_{\mu} a_\mu \sum_{m=-l}^{l} \delta q \, \rho_{l}^{m *}
\left(q_{\mu,l}^{m} + a_\mu \delta q \, \rho_{l}^{m} \right),
\label{Eq:Perturbed_part_of_action}
\end{align}
where $U(\hat{\bfr}\cdot\hat{\bfr}^{\prime})$ is the bare interaction,
expanded by
Eq.~\eqref{Eq:Effective_interaction_spherical_harmonics_expansion}
with coefficients $U_l$. Performing
the derivatives in Eq.~\eqref{Eq:Bare_second_derivative_free_energy} yields
\begin{align}
\frac{\partial^2 F [\delta q]}{\partial \delta q^2}\Bigg|_{\delta
q = 0} =& \sum_{l=0}^{\infty}\frac{8\pi}{2l+1}U_l \sum_{\mu,\nu} a_\mu
\delta_{\mu \nu} a_\nu \sum_{m=-l}^{l} \rho_{l}^{m *} \rho_{l}^{m}\nonumber\\
& - \sum_{l=0}^{\infty}\frac{4\pi}{2l+1} U_l \sum_{l^{\prime}=0}^{\infty}
\frac{4\pi}{2l^{\prime}+1} U_{l^{\prime}} \sum_{\mu,\nu} a_\mu a_\nu \nonumber\\
& \quad \times \sum_{m=-l}^{l} \sum_{m^{\prime}=-l^{\prime}}^{l^{\prime}}
\rho_{l}^{m *} \rho_{l^{\prime}}^{m^{\prime}} \left \langle
  q_{\mu,l}^{m} q_{\nu,l^{\prime}}^{m^{\prime} *}\right \rangle .
\label{Eq:Bare_second_derivative_free_energy_1}
\end{align}

We introduce the dielectric function $\epsilon_{\mu \nu,l}$ by
\begin{equation}
U^{\textnormal{eff}}_{\mu \nu, l} = \epsilon_{\mu \nu,l}^{-1} U_{l},
\label{Eq:Definition_dielectric_function}
\end{equation}
and by comparing
Eqs.~\eqref{Eq:Effective_second_derivative_free_energy} and
\eqref{Eq:Bare_second_derivative_free_energy_1}, the inverse dielectric
function is found to be
\begin{align}
\epsilon_{\mu \nu,l}^{-1} = \delta_{\mu \nu} -&
\left(\sum_{m=-l}^{l} \rho_{l}^{m *} \rho_{l}^{m}\right)^{-1}
\sum_{l^{\prime}=0}^{\infty} \frac{2\pi}{2l^{\prime}+1}
U_{l^{\prime}}\nonumber\\
&\times \sum_{m=-l}^{l} \sum_{m^{\prime}=-l^{\prime}}^{l^{\prime}} \rho_{l}^{m *} \rho_{l^{\prime}}^{m^{\prime}} \left \langle
  q_{\mu,l}^{m} q_{\nu,l^{\prime}}^{m^{\prime} *}\right \rangle.
\label{Eq:Inverse_dielectric_function}
\end{align}
Moreover, since the bare interaction is only dependent on the distance
between the charges, $U = U(\hat{\bfr}\cdot\hat{\bfr}^{\prime})$, we
have that $\langle q_{\mu,l}^{m} q_{\nu,l^{\prime}}^{m^{\prime}
    *}\rangle = \langle q_{\mu,l}^{m}  q_{\nu,l^{\prime}}^{m^{\prime}
    *}\rangle \delta_{ll^{\prime}}
  \delta_{mm^{\prime}}$, which yields
\begin{align}
\epsilon_{\mu \nu,l}^{-1} = \delta_{\mu \nu} -&
\left(\sum_{m=-l}^{l} \rho_{l}^{m *} \rho_{l}^{m}\right)^{-1}
\frac{2\pi}{2l+1} U_{l}\nonumber\\
&\times \sum_{m=-l}^{l} \rho_{l}^{m *} \rho_{l}^{m} \left \langle
  q_{\mu,l}^{m} q_{\nu,l}^{m *}\right \rangle.
\label{Eq:Inverse_dielectric_function_1}
\end{align}
Additionally, the property that the bare interaction is distance
dependent, only, yields an interaction $U_l$ that is independent of $m$. Hence, the correlator $\langle q_{\mu,l}^{m}
q_{\nu,l}^{m *}\rangle$ must be $m$ independent as well, $\langle q_{\mu,l}^{m}
q_{\nu,l}^{m *}\rangle = \langle q_{\mu,l}^{0}
q_{\nu,l}^{0}\rangle$. The dielectric function thus reads
\begin{equation}
\epsilon_{\mu \nu,l}^{-1} = \delta_{\mu \nu} - \frac{2\pi}{2l+1} U_l \left \langle q_{\mu,l}^{0} q_{\nu,l}^{0}\right \rangle.
\label{Eq:Inverse_dielectric_function_2}
\end{equation}

The dielectric constant $\epsilon_{\mu \nu}$ is now found in the long wavelength limit of
the dielectric function. On a spherical surface, this corresponds to
setting $l = 1$ in the dielectric function, i.e. $\epsilon_{\mu \nu} = \epsilon_{\mu \nu,1}$. Thus, the dielectric constant is
\begin{equation}
\epsilon_{\mu \nu}^{-1} = \delta_{\mu \nu} - \frac{2\pi}{3} U_1 \left \langle q_{\mu,1}^{0} q_{\nu,1}^{0}\right \rangle.
\label{Eq:Inverse_dielectric_constant}
\end{equation}

So far, only a few assumptions are made regarding the bare
interaction $U(\hat{\bfr}\cdot\hat{\bfr}^{\prime})$ and the charge
field $q_\mu(\theta, \phi)$. To apply
Eq. \eqref{Eq:Inverse_dielectric_constant} for the system under
consideration in this paper, we invoke
$U(\hat{\bfr}\cdot\hat{\bfr}^{\prime}) =
-\ln(1-\hat{\bfr}\cdot\hat{\bfr}^{\prime})$ to find $U_1 =
3/2$ by Eq. \eqref{Eq:Legendre_coefficients}. Moreover, the charge
field is modeled as point charges in a uniform background
\begin{equation}
q_\mu(\theta, \phi) = q^{\textnormal{BG}}_{\mu} + \sum_{i=1}^{N} e_{\mu,i}\frac{\delta(\theta-\theta_{i})\delta(\phi-\phi_{i})}{\sin \theta},
\label{Eq:Charge_field}
\end{equation}
where $q^{\textnormal{BG}}_{\mu} = -(\sum_{i} e_{\mu,i})/(4\pi)$ is the
uniform background ensuring charge neutrality for charges of type $\mu$,
$e_{\mu,i}$ is the charge of type $\mu$ in particle $i$ and the sum is
over all $N$ particles of the unperturbed system. Now, using
Eq. \eqref{Eq:Spherical_harmonics_coefficients}, the actual coefficient
of the charge field is found to be
\begin{equation}
q_{\mu,1}^{0} = \sqrt{\frac{3}{4\pi}}\frac{M_{\mu,z}}{R},
\label{Eq:Charge_field_coefficient}
\end{equation}
where $\bfM_\mu = \sum_{i=1}^{N} e_{\mu,i}\hat{\bfr}_i$ is the total
dipole moment for charges of type $\mu$. Finally, by inserting these results in
Eq. \eqref{Eq:Inverse_dielectric_constant}, the inverse dielectric constant is
found to be
\begin{equation}
\epsilon_{\mu \nu}^{-1} = \delta_{\mu \nu} - \frac{\pi}{A} \left \langle \bfM_\mu\cdot\bfM_\nu\right \rangle,
\label{Eq:Inverse_dielectric_constant_1}
\end{equation}
where $\langle M_{\mu,z}M_{\nu,z} \rangle = \langle
\bfM_\mu\cdot\bfM_\nu \rangle /3$ by assuming isotropy.

When there are test charges with multiple interactions, there
are multiple contributions to the change in free energy as seen in
Eq. \eqref{Eq:Free_energy_increase_by_effective_interaction}. To
account for all contributions to the increase in free energy, we
construct a generalized dielectric constant by
\begin{equation}
\epsilon_{(a_1, a_2, ...)}^{-1} = \sum_{\mu, \nu} a_\mu \epsilon_{\mu \nu}^{-1} a_\nu.
\label{Eq:General_inverse_dielectric_constant}
\end{equation}
Notice that even though there is no bare interaction
between charges of different type, there may be nonzero
cross terms in
Eq. \eqref{Eq:Free_energy_increase_by_effective_interaction}, as
charges of different type are constrained to be together within
the same particle.

\section{Fourth-order free energy derivative}
\label{App:Fourth_order_derivative}

In Ref.~\onlinecite{Minnhagen_PRB_2003} a method of verifying the
discontinuous character of the BKT-transition was introduced, by
examining a higher-order term in the free energy expansion in the
XY-model when the system is perturbed with an infinitesimal phase
twist. Similarly, in Ref.~\onlinecite{Borkje_PRB_2005}, the method
was applied in a two-dimensional logarithmic plasma. Here, we show
that the same idea also applies when we perturb a logarithmic Coulomb
plasma on a spherical surface with an infinitesimal test charge field
with multiple types of Coulomb interactions.

Consider a system with particles interacting with different charges of
multiple types, as previously described. We now \textit{choose} to perturb this
system with a neutral distribution of test charge of multiple
types, which has the form $\delta q_\mu (\theta) = a_\mu \delta q \cos(\theta)$,
i.e. a similar test particle density field as given in
Eq.~\eqref{Eq:Test_charge_field} but with $\rho_{1}^{0} = \sqrt{4\pi/3}$ being the
only nonzero coefficient in the spherical harmonics expansion. This is a convenient choice because
it corresponds to the most long-waved nonuniform test charge configuration on the surface of a
sphere, and hence, the prefactor of the second-order term in the free energy expansion
will be proportional to the inverse dielectric constant, as we will see
below.

The test charges yield a contribution to the potential energy as given
by the $l = 1$ and $m = 0$ term in
Eq.~\eqref{Eq:Perturbed_part_of_action},
\begin{equation}
V_1 [\delta q] = \frac{4\pi}{3} U_{1}
\sum_{\mu} a_\mu \delta q \, \rho_1^0 \left(q_{\mu,1}^{0} + a_\mu \delta q \rho_1^0 \right).
\label{Eq:Perturbed_part_of_action_l1}
\end{equation}

We now consider how the system responds to the test charges by a Taylor expansion
of the free energy in the test charge field around $\delta q = 0$,
\begin{align}
\Delta F[\delta q] =& \frac{\partial F[\delta q]}{\partial
  \delta q}\bigg|_{\delta q=0} \delta q + \frac{\partial^2
  F[\delta q]}{\partial\delta q^2}\bigg|_{\delta
  q=0}\frac{\delta q^2}{2!}\nonumber\\
&+ \frac{\partial^3 F[\delta
  q]}{\partial\delta q^3}\bigg|_{\delta q=0}\frac{\delta
  q^3}{3!} + \frac{\partial^4 F[\delta q]}{\partial\delta
  q^4}\bigg|_{\delta q=0}\frac{\delta q^4}{4!} + \ldots.
\label{Eq:Free_energy_expansion}
\end{align}
The change in the free energy $\Delta F[\delta q]$ must be invariant
to $\delta q_\mu (\theta) \rightarrow -\delta q_\mu (\theta)$, and
hence, all odd-order derivatives in Eq. \eqref{Eq:Free_energy_expansion} are
zero. From Appendix \ref{App:General_dielectric_constant} (see
Eqs. \eqref{Eq:Bare_second_derivative_free_energy_1}, \eqref{Eq:Inverse_dielectric_constant}
and \eqref{Eq:General_inverse_dielectric_constant}), the second-order free energy derivative
is found to be
\begin{equation}
\frac{\partial^2 F[\delta q]}{\partial\delta q^2}\bigg|_{\delta q=0} =
\frac{8\pi}{3} (\rho_1^0)^2 U_1 \epsilon_{(a_1, a_2, ...)}^{-1}.
\label{Eq:Appendix_Second_order_derivative}
\end{equation}
The fourth-order derivative is
\begin{align}
\frac{\partial^4 F[\delta q]}{\partial\delta q^4}\bigg|_{\delta q=0} =&
\, 3\left\langle \left(\frac{\partial V_1[\delta q]}{\partial\delta
      q}\bigg|_{\delta q=0}\right)^2\right\rangle^2\nonumber\\
&- \left\langle \left(\frac{\partial V_1[\delta q]}{\partial\delta q}\bigg|_{\delta q=0}\right)^4\right\rangle\nonumber\\
=& \left(\frac{4\pi}{3} \rho_1^0 U_1\right)^4
\sum_{\mu,\nu,\rho,\sigma}a_{\mu}a_{\nu}a_{\rho}a_{\sigma}\nonumber\\
\times \big[3\big\langle q_{\mu,1}^0q_{\nu,1}^0\big\rangle & \big\langle
 q_{\rho,1}^0q_{\sigma,1}^0\big\rangle-\big\langle
q_{\mu,1}^0q_{\nu,1}^0q_{\rho,1}^0q_{\sigma,1}^0\big\rangle\big].
\label{Eq:Appendix_Fourth_order_derivative}
\end{align}
where brackets denote a statistical average with respect to the unperturbed action. Inserting
Eqs. \eqref{Eq:Appendix_Second_order_derivative} and \eqref{Eq:Appendix_Fourth_order_derivative}
in Eq. \eqref{Eq:Free_energy_expansion} yields
\begin{align}
\Delta F[\delta q] = \frac{8\pi}{3} (\rho_1^0)^2 U_1 \Big[&\epsilon_{(a_1, a_2,
    ...)}^{-1}\frac{\delta q^2}{2!}\nonumber\\
&+ \gamma_{(a_1, a_2, ...)} \frac{\delta q^4}{4!} + \ldots \Big],
\label{Eq:Free_energy_expansion_1}
\end{align}
where
\begin{equation}
\gamma_{(a_1, a_2, ...)} = \sum_{\mu,\nu,\rho,\sigma}a_{\mu}a_{\nu}a_{\rho}a_{\sigma}\gamma_{\mu\nu\rho\sigma},
\label{Eq:General_fourth_order_term}
\end{equation}
and
\begin{align}
\gamma_{\mu\nu\rho\sigma} =
\left(\frac{4\pi}{3}U_1\right)^3\frac{(\rho_1^0)^2}{2}\big[&3\big\langle q_{\mu,1}^0q_{\nu,1}^0\big\rangle \big\langle
 q_{\rho,1}^0q_{\sigma,1}^0\big\rangle\nonumber\\
&-\big\langle q_{\mu,1}^0q_{\nu,1}^0q_{\rho,1}^0q_{\sigma,1}^0\big\rangle\big].
\label{Eq:Specific_fourth_order_term}
\end{align}

Now, inserting $\rho_{1}^{0} = \sqrt{4\pi/3}$ and assuming the charge field in Eq. \eqref{Eq:Charge_field} and a logarithmic bare interaction,
$U_1 = 3/2$, yields
\begin{align}
\gamma_{\mu\nu\rho\sigma} = \left(\frac{\pi}{R^2}\right)^2\big[&\left\langle
\bfM_{\mu}\bfM_{\nu}\right\rangle\left\langle  \bfM_{\rho}\bfM_{\sigma}\right\rangle\nonumber\\
&-3\left\langle M_{\mu,z}M_{\nu,z}M_{\rho,z}M_{\sigma,z}\right\rangle\big],
\label{Eq:Specific_fourth_order_term_1}
\end{align}
where $\langle M_{\mu,z}M_{\nu,z} \rangle = \langle
\bfM_\mu\cdot\bfM_\nu \rangle /3$ by assuming isotropy.

\subsection{Stability argument}

When $\delta q = 0$, the free energy of the system has a global minimum, and hence, the right-hand
side of Eq. \eqref{Eq:Free_energy_expansion_1} must be greater or equal to zero. Now, if
$\gamma_{(a_1, a_2, ...)}$ approaches a nonzero
negative value at the critical point in the thermodynamical limit, the general inverse
dielectric constant must simultaneously have a nonzero positive value for the ground
state to be stable. However, since $\epsilon_{(a_1, a_2, ...)}^{-1} = 0$
in the screening phase, it follows that $\epsilon_{(a_1, a_2, ...)}^{-1}$ must exhibit
a discontinuous jump at the critical point. Hence, investigation of $\gamma_{(a_1, a_2, ...)}$
may be used to verify a discontinuity in the inverse dielectric constant, which is a necessary
requirement for observing a BKT-transition.

\section{The finite-size scaling relation}
\label{App:Weber_Minnhagen}

The finite-size scaling relation of the BKT-transition has been used
throughout this article to verify the universal jump in
$\epsilon_{22}^{-1}$ and to provide estimates for the critical coupling
$Q_{2,\text{c}}^2$. Here, some details to the curve fitting procedure and the
goodness of fit measure are presented.

\subsection{Two free parameters}

Least-squares curve fitting of the Monte-Carlo results for
$\epsilon_{22}^{-1}$ to Eq.~\eqref{Eq:Finite_size_scaling} may be performed
with both $\epsilon_{22}^{-1}(\infty)$ and $C$ as free
parameters~\cite{Weber_PRB_1988, Minnhagen_PhysicaB_1988,
  Lee_PRB_1992, Gupta_PRB_1997}. If the
transition is of the BKT-type, a good fit to
Eq.~\eqref{Eq:Finite_size_scaling} should be obtained at the critical
point. In addition, when $\epsilon_{22}^{-1}(\infty)$ is free, no \textit{a priori} assumption on the
value of the universal jump is made, thus a resulting value of
$\epsilon_{22}^{-1}(\infty)$ that corresponds to the universal jump of
the BKT-transition should be obtained. However, with two free
parameters, higher quality of the Monte-Carlo statistics is
required to single out when they system is closely obeying the behavior of Eq.~\eqref{Eq:Finite_size_scaling}.

We have employed the Marquardt-Levenberg algorithm minimizing $\chi^2$
to the nonlinear fitting function in
Eq.~\eqref{Eq:Finite_size_scaling}. Specifically, $\chi^2$ is the sum
of squared weighted residuals,
\begin{equation}
\chi^2 = \sum_{i=1}^{n} \left(\frac{\epsilon_{22,N_{i}}^{-1}-\epsilon_{22}^{-1}(N_i)}{\sigma_{N_{i}}}\right)^2,
\label{Eq:Chi_squared}
\end{equation}
where $n$ is the number of system sizes $N_i$,
$\epsilon_{22,N_{i}}^{-1}$ is the value of the inverse dielectric
constant $\epsilon_{22}^{-1}$ obtained from the Monte-Carlo simulation
at system size $N_i$, and $\sigma_{N_{i}}$ is the corresponding error. For a
good fit, we expect the weight-normalized residuals,
$Y_i = (\epsilon_{22,N_{i}}^{-1}-\epsilon_{22}^{-1}(N_i))/\sigma_{N_{i}}$ to be
Gaussian-distributed with mean $\mu(Y_i) = 0$ and variance $\sigma^2 (Y_i) =
1$. Thus, to measure the goodness of the fit, we use the Anderson-Darling test
statistic $A^2$ for the data set $Y_i$ to arise from a normal
distribution with $\mu(Y_i) = 0$ and $\sigma^2 (Y_i) =1$:
\begin{equation}
A^2 = -n -\frac{1}{n}\sum_{i=1}^{n} \left(2i-1\right)\left\{\ln[\Phi(Y_i)]+\ln[\Phi(Y_{n+1-i})]\right\},
\label{Eq:Anderson_Darling}
\end{equation}
where $\Phi(Y)$ is the standard normal cumulative distribution
function and where the data set $Y_i$ is ordered from low to high values.
A smaller value of $A^2$ essentially means a better fit
between the data and the fit function.

\begin{figure}[tbp]%%[htbp]
\includegraphics[width=\columnwidth]{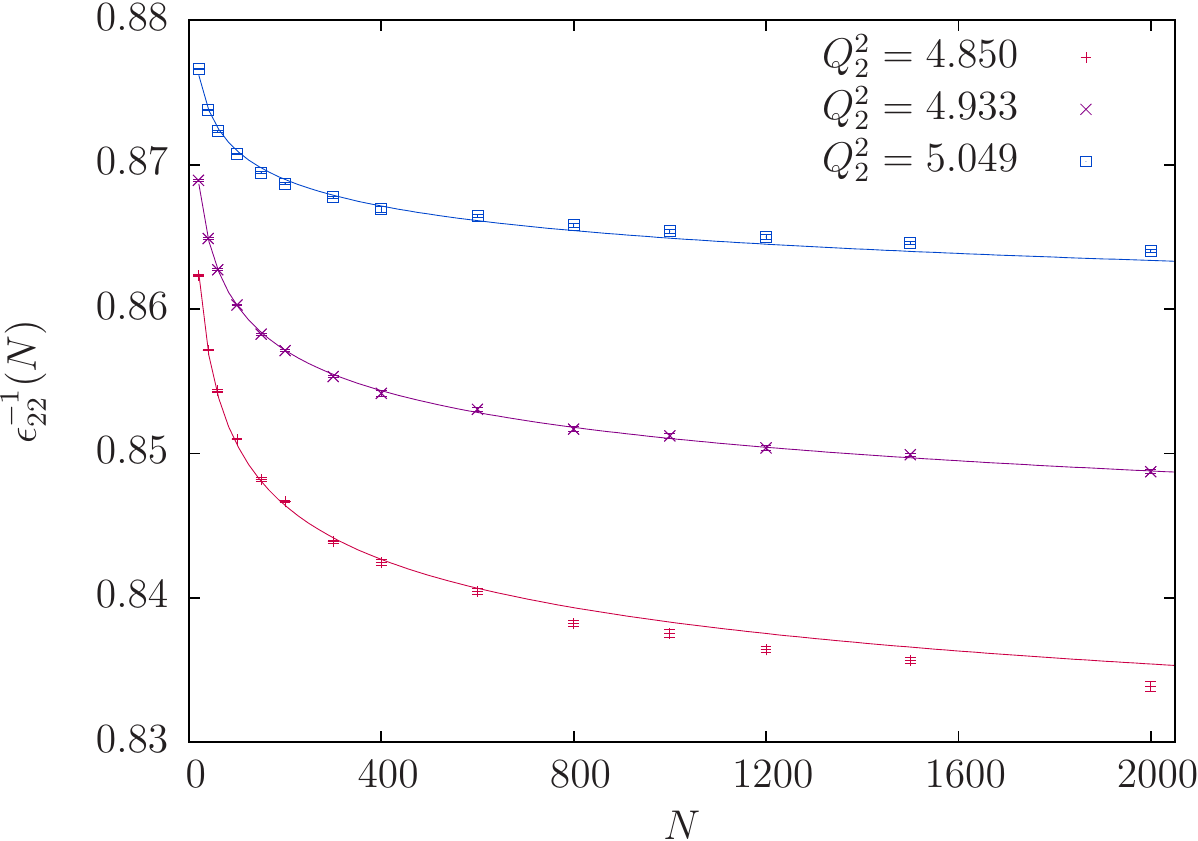}
\caption{(Color online) Plot of the size-dependence in the inverse dielectric constant
  $\epsilon_{22}^{-1}(N)$ for fourteen different system sizes in the range $20
  \leq N \leq 2000$ at three different values of the coupling
  $Q_2^2$. The best fit according to the fit function in
  Eq. \eqref{Eq:Finite_size_scaling} with two free parameters, is
  given as the corresponding solid line in all three cases. The packing
  fraction is $\eta = 2\cdot 10^{-3}$ and $Q_1 = 0$.}
\label{Fig:Curvefit_finite_size}
\end{figure}

\begin{figure}[t!]%%[htbp]
\includegraphics[width=\columnwidth]{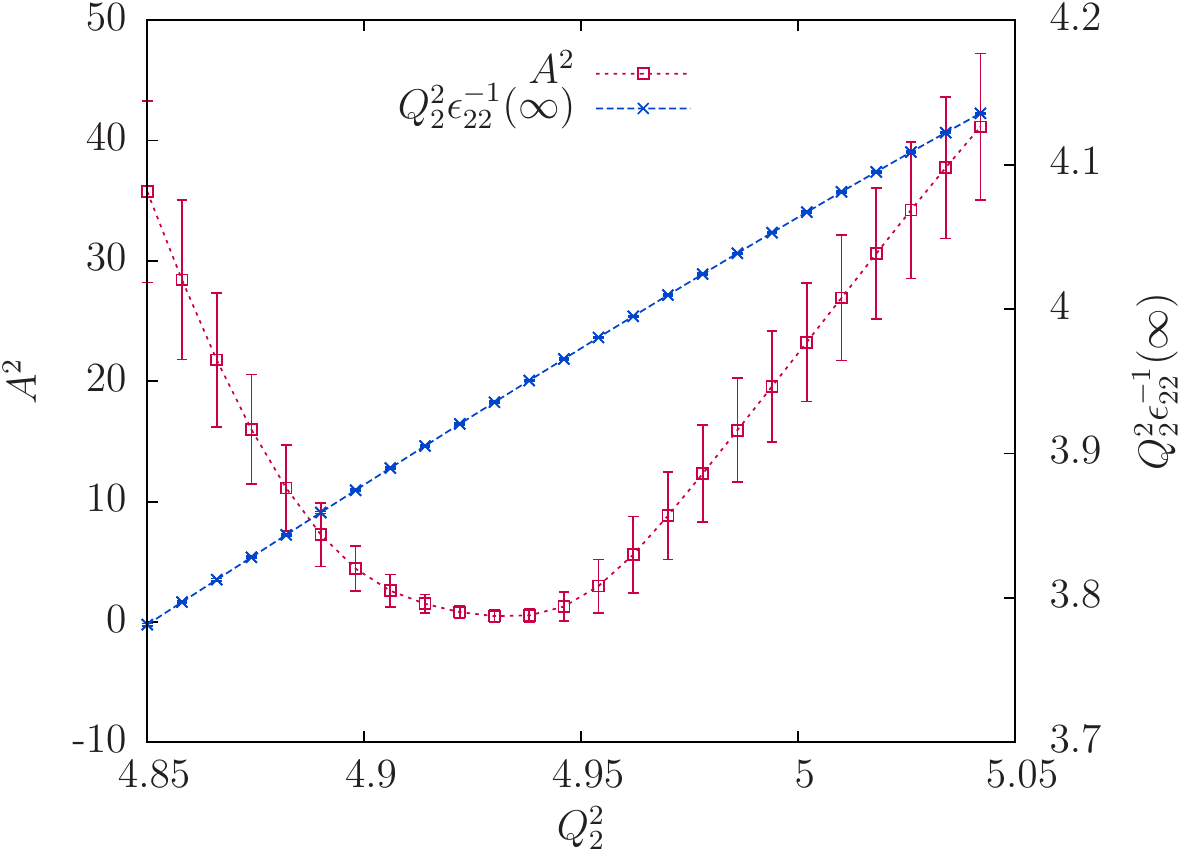}
\caption{(Color online) Plot of the goodness of fit parameter $A^2$ and the
  corresponding free parameter $\epsilon_{2}^{-1}(\infty)$ obtained when
  curve fitting to the critical finite-size relation given in
  Eq.~\eqref{Eq:Finite_size_scaling}. The results are given as a
  function of $Q_2^2$. System sizes $N$, and $\eta$ and $Q_1$ are the same
  as in Fig.~\ref{Fig:Curvefit_finite_size}. Error estimates are
  obtained by the Jackknife method.}
\label{Fig:Curvefit}
\end{figure}

To illustrate the method, Monte-Carlo results for $\epsilon_{22}^{-1}$ at
fourteen different system sizes and the corresponding curve-fit according to
Eq.~\eqref{Eq:Finite_size_scaling} are given in Fig.~\ref{Fig:Curvefit_finite_size} for three different
values of $Q_2^2$. Here, $\eta = 2\cdot 10^{-3}$ and $Q_1 = 0$. Clearly,
at $Q_2^2 = 4.933$, the fit between the data and the fit function is
better than for the two other cases. Moreover, in Fig.~\ref{Fig:Curvefit} the corresponding results for the goodness of fit
parameter as well as the results for
the parameter $\epsilon_{22}^{-1}(\infty)$ as a function of $Q_2^2$ are
shown. Indeed, the minimum in $A^2$ indicates a critical region where
the data seem to follow the logarithmic finite size scaling of
$\epsilon_{22}^{-1}$ given in Eq.~\eqref{Eq:Finite_size_scaling}. Also note
that this region coincides with a value of $Q_2^2 \epsilon_{22}^{-1} (\infty)$
close to the universal jump value of 4. With the minimum of $A^2$ as a
measure of the critical point and with error estimates obtained by the
Jackknife method, we find that $Q_{2,\text{c}}^2 = 4.933 \pm 0.012$
and that $Q_{2,\text{c}}^2\epsilon_{22}^{-1}(\infty) = 3.941 \pm 0.023$, less than
2\% off the universal number. The results in Figs.~\ref{Fig:Critical_QQ_Weber_Minnhagen_method}
and \ref{Fig:Universal_jump_Weber_Minnhagen_method} are found by repeating
this procedure for different values of $\eta$ and $Q_1$.

\subsection{One free parameter}
The procedure described in detail above with two free parameters, may be
performed with a fixed value of $\epsilon_{22}^{-1}(\infty) =
4Q_{2,\text{c}}^2$ and with $C$ as the only free parameter. If the
transition is of the BKT-type, a good fit to
Eq.~\eqref{Eq:Finite_size_scaling} should be obtained at the critical
point. This is a rather well-used method to determine the critical point of a
BKT-transition~\cite{Weber_PRB_1988, Lidmar_PRB_1997, Bonnes_PRL_2011,
Kuroyanagi_JLowTempPhys_2011}. With only one free parameter, $Q_{2,\text{c}}^2$
will be determined with greater accuracy compared to the case when
there are two free parameters.

\subsection{Remarks}
Refs.~\onlinecite{Lee_PRB_1992} and \onlinecite{Gupta_PRB_1997} used
$\chi^2$ as a goodness of fit parameter. We also tried this, and the results
for the critical coupling as well as the corresponding parameter $\epsilon_{22}^{-1}(\infty)$
were consistent with $A^2$ results within statistical uncertainty. However, we
found that error estimates were clearly underestimated with $\chi^2$, probably
due to over-fitting.

The parameter $C$ in the finite-size scaling relation
(Eq. \eqref{Eq:Finite_size_scaling}) is density
dependent~\cite{Olsson_PhysicaScripta_1991}. Specifically, $C$
increases when $\eta$ decreases. Hence, at the critical point, the
finite-size scaling slows down when $\eta$ is lowered. Therefore,
larger systems $N$ or better statistics are required to resolve the
critical scaling when $\eta$ is small. In particular, curve fitting to
Eq.~\eqref{Eq:Finite_size_scaling} was also performed for $\eta =
5\cdot 10^{-5}$ in addition to the densities presented in
Figs.~\ref{Fig:Critical_QQ_Weber_Minnhagen_method} and
\ref{Fig:Universal_jump_Weber_Minnhagen_method}. However, in this case
the statistics were not good enough to resolve a clear minimum in
$A^2$. Also note that there are higher-order
corrections\cite{Olsson_PhysicaScripta_1991} to
Eq.~\eqref{Eq:Finite_size_scaling} that are not taken into account in this
work.

\end{document}